\documentclass[aip,amsmath,amssymb,jcp,12pt
]{revtex4-1}
\usepackage[utf8]{inputenc}
\usepackage{soul}

\usepackage{amsfonts}
\usepackage{amsmath}
\usepackage{amssymb}
\usepackage{amsthm}
\usepackage{appendix}
\usepackage{bbm}
\usepackage{chemformula}
\usepackage{enumerate}
\usepackage[T1]{fontenc}
\usepackage{datetime}
\usepackage{dcolumn} 
\usepackage{epigraph}
\usepackage[T1]{fontenc}
\usepackage{graphicx} 
\usepackage{hyperref}
\hypersetup{
    colorlinks=true,
    linkcolor=blue,
    filecolor=magenta,
    urlcolor=cyan,
}
\usepackage{cleveref}
\usepackage{tabularx}
\usepackage{xcolor}
\usepackage{yfonts}

\newcommand{\bfr}{{\bf r}}

\newcommand{\R}{\mathbb{R}}

\newcommand{\alert}[1]{#1} 

\begin{document}

\title{The effect of uncertainty on building blocks in molecules}

\author{Anthony Scemama}
\address{Laboratoire de Chimie et Physique Quantiques (UMR
  5626), Universit\'e de Toulouse, CNRS, UPS, France \\
  scemama@irsamc.ups-tlse.fr}
\author{Andreas Savin}
\address{CNRS and Sorbonne Universit\'e, Laboratoire de Chimie Th\'eorique (LCT),\\
  F-75005 Paris, France\\
  andreas.savin@lct.jussieu.fr}

\begin{abstract}
Probabilities to find a chosen number of electrons in flexible domains of space
are calculated for highly correlated wave functions.  Quantum mechanics can
produce higher probabilities for chemically relevant arrangements of electrons
in these regions.  However, the probability to have a given arrangement, e.g.,
that corresponding to chemical formulas (bonds or atoms), is low although
being often maximal.  Like in valence bond theory, it is useful to consider
alternative distributions of the electrons.  Exchanges of electrons should be
considered not only between atoms, but also between other types of regions,
like those attributed to lone pairs.  It is useful to have definitions flexible
enough to allow the user to find the reference representations he considers
most relevant.  We tentatively suggest a tool (the effective number of parties)
to help one make the choice.

\end{abstract}

\keywords{atoms, chemical bond, effective number of electrons, probabilities}

\maketitle

\tableofcontents

\centerline{\today \hspace{2pt} at \currenttime}

\section{Introduction}

Building blocks are often used to understand, and even solve problems.
For two centuries, the atom has been considered the most important building block of molecules.
Associating the words ``atomic'' and ``hypothesis'' (as was done in the 19th century to deny their existence) is -- in popular view --  ridiculous: we know that atoms exist.
However, in quantum chemistry we perform our calculations using nuclei and electrons:
the Coulomb forces exerted by the nuclei, and the requirement of antisymmetry shape the wave function.
Due to these factors, the electrons do not behave independently from one another, they do not roam freely through space.
This leaves room for searching quantities that correspond to building blocks, in spite of difficulties that arise when doing it:
\begin{enumerate}
    \item Quantum calculations cover a wider field.
    There is room for aspects not covered by traditional concepts.
    For example, Lewis wrote that, for him, the three-electron bond was unthinkable.~\cite{Lew-33}
    \item The association between quantities derived from quantum calculations and traditional building blocks is not unique.
    \alert{For example, e}ven a rigorous choice of a definition of an atom in a molecule remains nothing but one of possible choices
    \alert{; one may prefer atoms as defined by Bader,~\cite{Bad-90} or ones from the Hirshfeld family.~\cite{HeiAyeVerVinVohBul-17}}
    \item Although the attractive forces of the nuclei confine electrons to the vicinity of nuclei, and the Pauli exclusion principle drives electrons to divide space among themselves, the separation into building blocks (atoms or electron pairs) is not perfect.
    Despite the constraints with confining effect, electrons can escape the regions of space we associate with the building blocks.
    \alert{Furthermore, the building blocks may not be uniquely determined.}

    \alert{Let us consider, as an example, the one-dimensional Kronig-Penney model.~\cite{Sav-21}
     A parameter describes barriers separating regions where pairs of electrons are localized.
     As electrons do not interact, we may as well consider as well half as many electrons, all of the same spin.
     When the barrier is infinite, we are sure to have one electron in its domain.
     When the barrier is reduced, electrons start escaping the domain between the barriers and the probability to find one and only one electron (of given spin) in it is reduced, but is still significant (around 2/3), due to the Pauli principle.
     Interestingly, another effect can be noticed.
     The domains are not so clear anymore.
     In the limit of zero barrier, the translation of the domain has no effect on the probability to have an electron in the domain (while it is significant when the barrier is large).
     The Pauli principle continues to act, even within delocalization.
     }
    \item The choice of the blocks we look at is not obvious.
    For example, is it not better to look for ions in an ionic crystal than for atoms?
    \alert{The spatial region defined by choosing the ``best'' atom is different from the one defined to find the ``best'' ion.
    Thus, the characterization of ``chemically'' equivalent quantities (for example, corresponding charges) differ between the two regions.}
\end{enumerate}

In this paper we focus on the last two points above.
We can quantify the uncertainty of electrons to be within a given region of space by calculating probabilities.
We can go even further, and ask about the effect a given number of electrons in one region of space has on the  number of electrons in another region of space.
Furthermore, we choose to give sufficient flexibility to the definitions to allow comparisons between different choices of the building blocks, for example having a number of electrons in a region corresponding to the neutral atom, or to one of its ions.

We would like to avoid -- as much as possible -- effects due to artefacts due to the level of computation.
Although our conclusions could be reached with less elaborate wave functions, we base our numerical results on quite accurate wave functions (linear combinations of up to millions of determinants, see details in appendix~\ref{app:technical}).

\alert{We would like to stress that certain aspects of our paper are not new.
Computing probabilities in spatial domains can be found already in older papers, like Refs.~\onlinecite{Sav-02, Sce-05, ChaFueSav-03}.
In the first two papers the domains were deformed to maximize probabilities, in the third paper, the probabilities were computed for commonly used spatial domains.
Probabilities that can be computed with elaborate wave functions also exist already, see, e.g.,  Refs.~\onlinecite{PenFraBla-c-07, SceCafSav-07, Reu-Luc-20}.
Joint and conditional probabilities were already presented in the context of partitioning.~\cite{Dau-53,GalCarLodCanSav-05,PenFra-19}
Comparisons with valence bond calculations exist already in the literature, for example, in Refs.~\onlinecite{PenFraBla-c-07,turek_2017,hiberty_2018,Reu-Luc-20}.
}

\section{Probabilistic measures}
\subsection{The probability to have $n$ electrons present in a spatial region}

The number of electrons in a spatial region is a \alert{discrete} random variable
\alert{that can take the values $0, 1, 2, \dots N$, and
whose average is known as the population.
When we select a region of space, there can be only an integer number
($n \in [0,N]$) of electrons inside the region.
However, the \emph{average} number of electrons in a region of space,
obtained by integrating the density over this region,
can be a fractional number.
Let us discuss these aspects in more detail for an arbitrary wave function $\Psi$.}
A Monte Carlo sampling \alert{of the electron positions} according to \alert{the 3$N$-dimensional density} $|\Psi|^2$ produces a large set of $M$ configurations (sets of positions for the $N$ electrons of the system, $\{\bfr_1, \bfr_2, \dots \bfr_N\}$; note that we use in this paper the term ``configuration'' in the sense of Monte Carlo, and not in the meaning of a linear combination of Slater determinants).
We estimate from the set of configurations the probability of a chosen number of electrons, $n$, $0 \le n \le N$ to be in a selected region of space, $\Omega$.
\begin{equation}
    P_\Omega(n)= \frac{m_\Omega(n)}{M} ,
\end{equation}
where $m_\Omega(n)$ is the number of configurations $\{\bfr_1, \bfr_2, \dots \bfr_N\}$ for which $n$ of the $\bfr_i$ are within $\Omega$.
Another way to describe $P_\Omega(n)$ is to use the indicator function, $\mathbbm{1}_\Omega(n)$.
It is equal to one, if $n$ of the $\bfr_i$ are within $\Omega$, and equal to zero otherwise.
Its expectation value gives the probability
\begin{equation}
    P_\Omega(n) = E(\mathbbm{1}_\Omega(n))
\end{equation}

Note that, as $\mathbbm{1}^2 = \mathbbm{1}$, the variance is easily obtained as
\begin{equation}
\label{eq:sigma}
    \sigma_\Omega(n)^2 = P_\Omega(n) - P_\Omega(n)^2
\end{equation}
so that we can estimate the error of $P_\Omega(n)$ from the variance of the mean, $\sigma_\Omega(n)/\sqrt{M}$,
using that $\sqrt{P_\Omega(n) - P_\Omega(n)^2}$ is between 0 and 1/2.
This shows that using $M$ of the order of $10^4$ already gives reasonable estimates of the probability, which in our case is not required with more than two decimals.
The results shown in is paper are produced using sets of the order of $10^6$ Monte Carlo configurations.
When not impeding clarity, we omit to indicate explicitly $n$ or $\Omega$; for example we will write
$P_\Omega \equiv P_\Omega(n)$, or $P(n) \equiv P_\Omega(n)$.

We can also obtain the population of the region $\Omega$,
\begin{equation}
\label{eq:pop}
    Q_\Omega = \sum_{n=0}^N n \, P_\Omega(n)
\end{equation}
It gives the average number of electrons in $\Omega$.
The same number would be obtained by integrating the electron density over the region $\Omega$.
\alert{To see it, let us start with the definition of the density,
\[ \rho(\bfr) =  \left\langle \Psi \left| \sum_{i=1}^N \delta(\bfr-\bfr_i) \right| \Psi \right\rangle \]
Integrating it over a spatial domain $\Omega$ we get the population of $\Omega$:
\[ \int_\Omega \rho(\bfr)d \bfr =  \left\langle \Psi \left| \sum_{i=1}^N \int_\Omega \delta(\bfr-\bfr_i) d\bfr \right| \Psi \right\rangle . \]
The operator appearing on the right hand side counts the number of electrons in $\Omega$ because
\[ \int_\Omega \delta(\bfr-\bfr_i) d\bfr =
\left\{
\begin{array}{ll}
      1 & \bfr_i \in \Omega \\
      0 & \bfr_i \notin \Omega
\end{array}
\right. \]
By taking the expectation value, we obtain the mean number of electrons in $\Omega$, as in Eq.~\ref{eq:pop}.
}

Other than in earlier papers (see, e.g., reference~\onlinecite{SceCafSav-07}), we are interested in considering more than one spatial region.
Let us use for the spatial regions the symbols $A, B, \dots$ (instead of $\Omega_A, \Omega_B$,\dots).
The joint probability of having $n_A$ electrons in $A$ and $n_B$ electrons in $B$ is
\begin{equation}
    P_{A \cap B}(n_A, n_B)=\frac{m_{A \cap B}(n_A,n_B)}{M}
\end{equation}
where $m_{A \cap B}(n_A,n_B)$ is the number of configurations for which $n_A$ electrons are in $A$ and $n_B$ electrons in $B$.
We can also write
\begin{equation}
    P_{A \cap B}(n_A, n_B)= E(\mathbbm{1}_A(n_A)\,\mathbbm{1}_B(n_B))
\end{equation}
Note that $P_A(n_A)$ can be seen as a marginal,
\begin{equation}
    P_A(n_A) = \sum_{n_B=0}^N P_{A \cap B}(n_A,n_B).
\end{equation}
Of course, $P_{A \cap B}$ can be extended to more than two regions, to the probability to have $n_A$ electrons in $A$, $n_B$ electrons in $B$, $n_C$ electrons in $C$, etc.:
$P_{A \cap B \cap C \dots}$.
If we have a partitioning of space, we will denote it by  $n_A|n_B|n_C \dots$

One might naively expect that the probability of $n_A$ electrons to be in $A$ and $n_B$ electrons to be in $B$, $P_{A \cap B}$, is  given by $P_A\, P_B$.
In general, this is not true, because our random variables are not necessarily independent.
A trivial example for this shows up when $B$ is the part of space not covered by $A$, $\R^3 \backslash A$, and $n_B=N-n$.
(One can see it as the situation $n_A|N-n_A$.)
In this case, $P_B=P_A$ and $P_{A \cap B} = P_A \ne P_A P_B = P_A^2$.
A way to detect the dependence of random variables is given by conditional probabilities, the probability of $n_A$ electrons to be in $A$ given that $n_B$ electrons are in $B$,
\begin{equation}
 \label{eq:cond}
    P_{A|B}= \frac{P_{A \cap B}}{P_B}.
\end{equation}
When $P_{A|B}=P_A$ the variables are independent.
This reminds us that, in general, $P_{A|B} \ne P_{B|A}$, although in daily life we easily mix them up.
Note that $P_{A|B}$ can be larger or smaller than $P_A$.
For example, for $A=\R^3 \backslash B$,  $P_{A|B}(n_A=N-n_B, n_B) =1$, $P_{A|B}(n_A \ne N-n_B, n_B) =0$, no matter what $P_A \in (0,1)$ is.

\section{Choice of the spatial region}

\subsection{Sharp boundaries}
In this paper we choose regions having sharp boundaries: an electron either is, or is not in $\Omega$.
This is close to what is done, for example, in loge theory,~\cite{Dau-53} or in the Quantum Theory of Atoms and Molecules (QTAIM).~\cite{Bad-90}
It would not be difficult to extend the definition of the region by having fuzzy boundaries (simply by replacing the indicator function by some other function).
This could bring us closer to other ways of analyzing the wave function, for example in the Mulliken population analysis, or in Valence Bond Theory (where fuzziness is due to considering atomic orbitals).
However, using fuzzy boundaries introduces a supplementary arbitrariness.
Furthermore, contrary to intuition, fuzzy boundaries can have an equalizing effect on the probabilities (see, appendix~\ref{sec:smooth}).
This is opposed to our aim to have one of the probabilities to be significantly larger than the other ones.

In principle, the regions $A$ and $B$ can overlap.
This appears, for example, when producing maximum probability domains,~\cite{SceCafSav-07} or single electron densities.\cite{Luc-11}
Furthermore,  the spatial regions may not cover all of $\R^3$.
The latter case reminds of the population analysis first promoted by Davidson.~\cite{Dav-67}
He argues that population analysis has a meaning only in a minimal basis.
A part of the density is not recovered this way, thus
Roby extended it~\cite{Rob-74} to take into account the missing contributions by defining two-, and multi-center contributions.
Such extensions are also possible for our type of discussion, but we leave it for further studies.

\subsection{Partial optimization of the spatial domain}

\label{app:shapes}
\begin{figure}[h!]
    \centering
    \includegraphics[width=0.9 \textwidth]{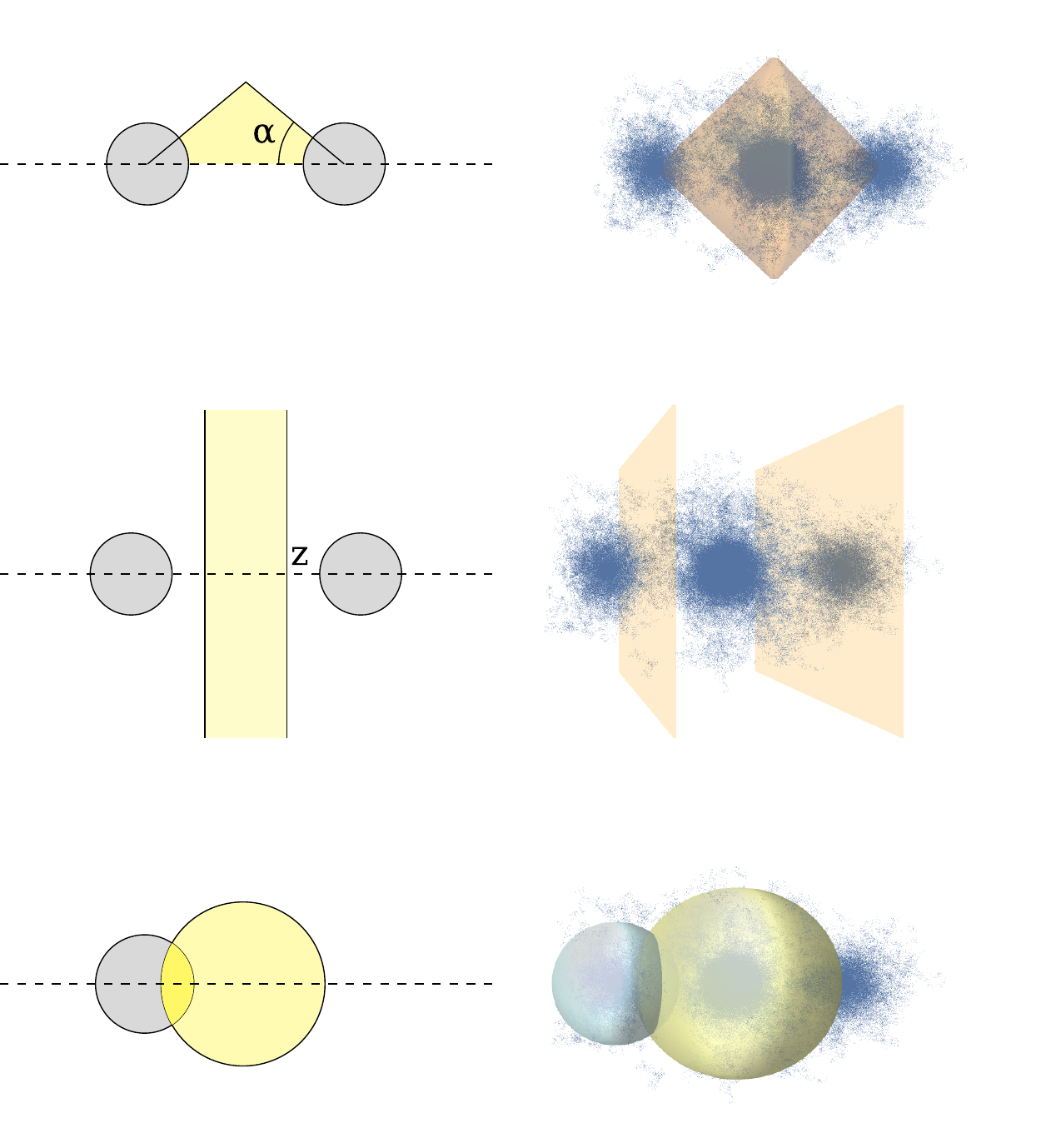}
    \caption{On the left, schematic representation of the geometrical
    shapes used for the separation of the domains. $2\alpha$ is the aperture of the cone, and $\pm z$
    are the positions of the planes separating the domains, and the dashed curves are the rotation axes.
   On the right side of the figure, three-dimensional representations of the domains are displayed, together with $\sim 10^4$  Monte Carlo samples in \ch{XeF2}. }
    \label{fig:sketch}
\end{figure}

For the present study, defining the spatial region by maximizing the probability to have a given number of electrons in it~\cite{Sav-02} seems appropriate.

However, in the present paper we perform only partial optimizations, for three reasons.
First, the points we try to make with our paper can be made also with a partial optimization (or just considering symmetry).
We try to define regions having a large (even if not maximum) probability.
Next, borders of the regions considered may contain a small number of configurations, and thus shape optimization needs a more careful treatment.
Finally, we would like our results to be easily reproducible without any important programming effort (as the optimization of the spatial domain requires).
\alert{
From now on, we qualify as ``optimal'' and ``best'' the region that maximizes the desired probability by varying the parameters defining the domain.
Note that the difference between the partially and the fully optimized spatial domain are transmitted only to second order to the probabilities.
}

For the spatial regions $\Omega$, we consider simple geometrical shapes that we optimize.
As a first example, let us consider the atomic cores.
Although they can be polarized by the molecular environment, we find that they are close to spheres centered on the nuclei.
We adjust the radius to maximize the probability of $n_{\text{core}}$ electrons to be in the sphere ($n_{\text{core}}$ is the number of electrons commonly assumed to be in the core, e.g., two for C and for F).

As the common image of atoms is spherical, let us first try to construct spheres around nuclei such that the probability of finding $n$ electrons is maximal.
Let us take, as an example, the \ch{XeF2} molecule (see, Fig.~\ref{fig:sketch}, panel bottom right).
The dots show $10^4$ configurations (each of 72 electron positions).
The sphere in the center maximizes the probability to find 54 electrons around the Xe nucleus.
It has a radius of $\approx 3.1$~bohr.
The other sphere, with a radius of $\approx 2.0$~bohr, maximizes the probability to find 9 electrons around one of the F nuclei.
It is clear that the spheres significantly overlap (the Xe-F distance is $\approx 1.4$~bohr smaller than the sum of the radii of the spheres).
Furthermore, as one can see in Fig.~\ref{fig:sketch}, a non-negligible number of  electron positions around a given atom are not taken in consideration for computing the probability.
This can be explained in the following way.
As the radius of the sphere around nucleus $A$ increases from 0, the probability to have $n>0$ electrons also increases.
Beyond a certain value of the radius, $P_A(n)$ continues to increase by including electrons that by there positions would normally be attributed to another atom, although -- for a flexible shape -- it would be possible to extend without doing it by deforming the sphere.
For $n<N$, after a certain radius, the $P_A(n)$ starts decreasing, as $P_A(n+1), P_A(n+2), \dots P_A(N)$ increase.
In particular, $P_A(N)=1$ for an infinite radius.
As the probabilities sum to one for a certain radius, the sphere stops increasing leaving out a number of  electron positions that would have been included allowing deformations of the sphere.
In the most accurate calculations of the domains maximizing the probability to find a pair of electrons describing bonds,~\cite{BraDalDapFre-20} it has been often found that they have shapes of cones with the apex on nuclei.
Of course, the core regions must be excluded.
A schematic drawing is shown in Fig.~\ref{fig:sketch}, top left panel.
We construct such regions by varying the aperture ($2 \alpha$) to maximize the probability to find the chosen numbers of electrons in it.
When the aperture is $\pi$, we have planes perpendicular to the rotation axis.
We found it convenient to displace the position of such planes, in particular when trying to define ``atoms'', e.g.
in the \ch{XeF2} molecule.
In fact, this corresponds to using weighted Voronoi cells.
For example, in \ch{XeF2}, oriented along the $z$-axis with the Xe atom at the origin, we find the value of $z$ for which the probability of having $n=54$ electrons between the two planes $\pm z$ is maximal~(see Fig.~\ref{fig:sketch}, middle panels).
Note that by using cones or planes, we have space-filling regions, and the problem of leaving out configurations (present for spheres) does not show up (see Fig.~\ref{fig:sketch},  panels top and middle right).

\subsection{Ambiguity in the choice of the spatial region}

The choice for a region depends on the object of study.
For example, in an ionic crystal, we may be more interested in finding ions than in finding atoms.
For some methods, one can explicitly formulate the requirement.
For finding maximum probability domains~\cite{Sav-02} the user specifies the number of electrons attributed to it.~\footnote{Multiple solutions may exist.
For example, for $n=2$, the region may correspond to the core of the O atom, to one of the lone pairs, or to the OH bond.}
This is the path we choose in this paper.
For other methods, it shows up in the result.
For example, in ionic crystals, the QTAIM ``atoms'' are closer to ions, see, e.g., refs.~\onlinecite{PenCosLua-98,CauSav-11}.)

\section{Fluctuations}

\subsection{Ehrenfest urn model}

In order to have a reference case, let us consider the Ehrenfest urn model (constructed to describe the fluctuations at thermal equilibrium).~\cite{EhrEhr-06}
$N$ balls (electrons for us) are in distributed in two urns (spatial domains for us), $A$ and $B$.
One ball is randomly picked from any of the urns (naturally, more likely with the urn with more balls) and put in the other urn.
As time evolves, the number of balls in the urns stabilizes to a distribution~\cite{Kac-47}, that -- for even $N$ -- has maximum at $N/2$ equal to $2^{-N} N!/[(N/2)!]^2$; the number of balls in an urn follows a binomial distribution.
The ratio between the probability to have $N/2$ balls in
the urn and that to have $N/2 \pm 1$ is $1+2/N$.
For example, for $N=4$ (two electron pairs), we would have $P_\Omega(n=2)=0.375$ and $P_\Omega(n=1)=P_\Omega(n=3)=0.25$.
As $N$ increases, not only $P(N/2)$ decreases, but $P(N/2 + 1)$ and $P(N/2-1)$ get closer to it.
For example, if we consider $N=14$ (two F atoms, valence shells), $P(7) \approx 0.21$ and $P(6) \approx 0.19$.

The Ehrenfest model can be extended to more than two urns,~\cite{Sie-49} leading to a multinomial distribution under assumptions similar to those made for two urns (see, e.g., ref.~\onlinecite{KarMcG-65}).
Let us consider, for example, the case of having six balls in three urns.
The probability to have exactly two balls in {\em one} of the urns is  $80/243 \approx 0.33$.
That of having two balls in {\em each} of the urns is lower, $30/243\approx0.12$.

\subsection{Orbital nodal structure and localization: two particles in a box}

Let us start with a simple analytic example.
We consider two non-interacting particles in a one-dimensional box: $0 \le x \le 1$.
The eigenfunctions are
\begin{equation}
    \phi_k(x) = \sqrt{2} \sin (k \, \alert{\pi} \, x), \; \text{with }k = 1, 2, \dots.
\end{equation}
We occupy $\phi_1$ with one electron and put the other electron on $\phi_k$,
\alert{ and construct Slater determinants using $\phi_1$, and $\phi_k$.}
A different spin is important only when both electrons are in $\phi_1$. Otherwise, it does not matter: there is no interaction in our model.
We ask about the probability to have one electron in one half of the box, and the other electron in the other half of the box.
For all odd $k$ it is equal to 1/2.
For even $k$, it is given by
\[ \frac{1}{2} + \frac{8}{\pi^2} \left(\frac{k}{k^2 - 1}\right)^2 \]

It is no surprise that for $k=1$ we obtain the urn result: the particles are independent.
The  probability is maximal (0.86) when $k=2$: the two electrons are now separated to a higher degree.
If the spins of the electrons were parallel, we would say this is due to the Pauli principle.
However, this holds also for antiparallel spins; we can attribute it to the orthogonality of the orbitals.
For odd $k \ne 1$, this principle works, too, but in each of the half-boxes: it does not help separating the electrons each into a half-box.

\begin{figure}[h!]
    \centering
    \includegraphics{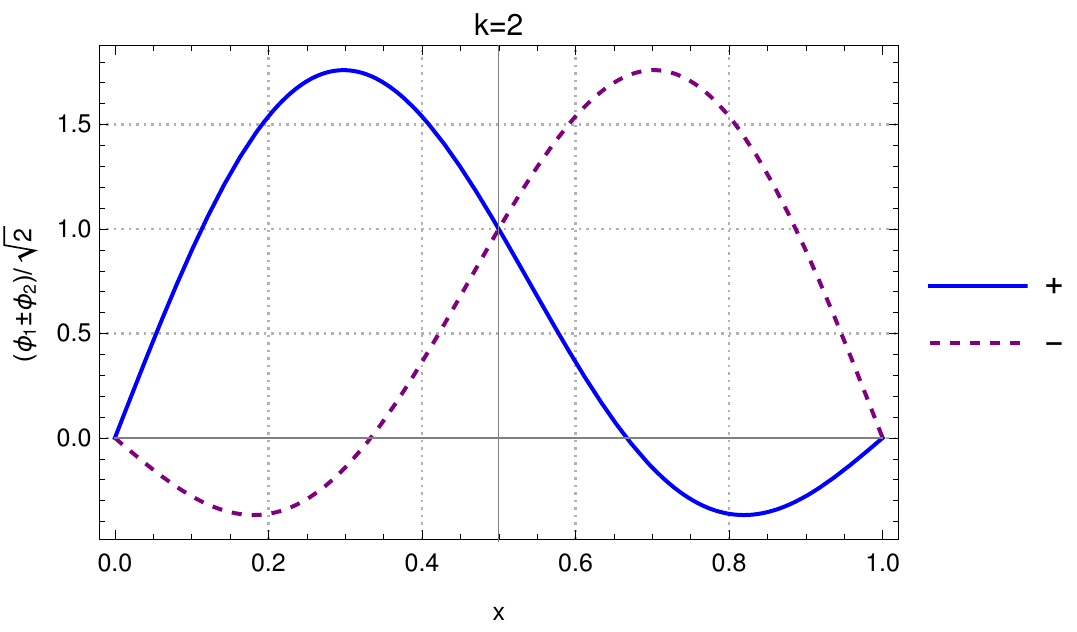} \\
    \includegraphics{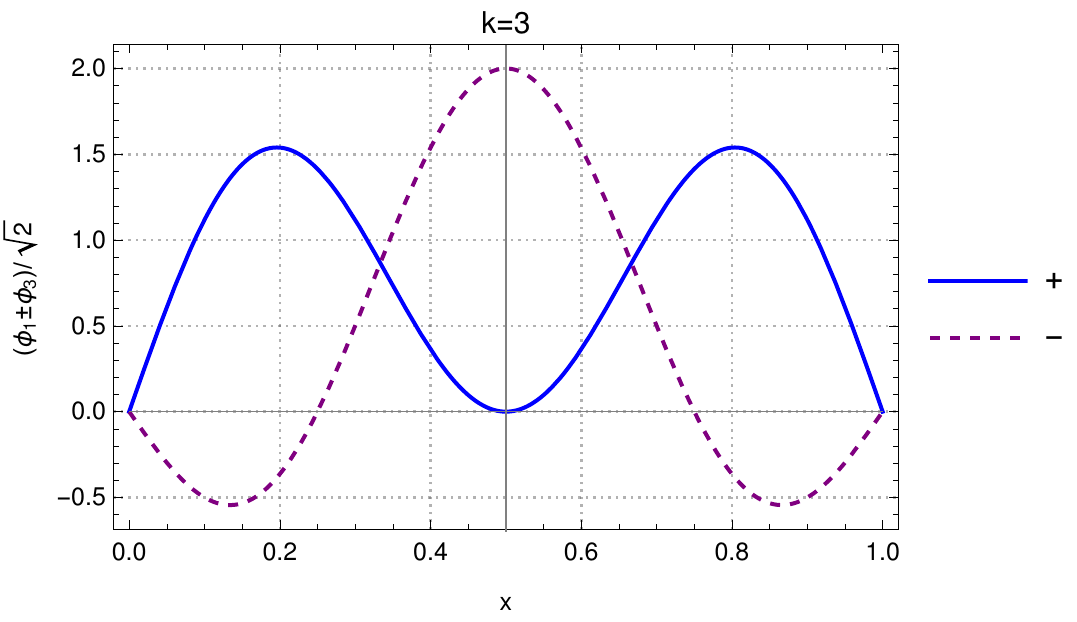} \\
    \includegraphics{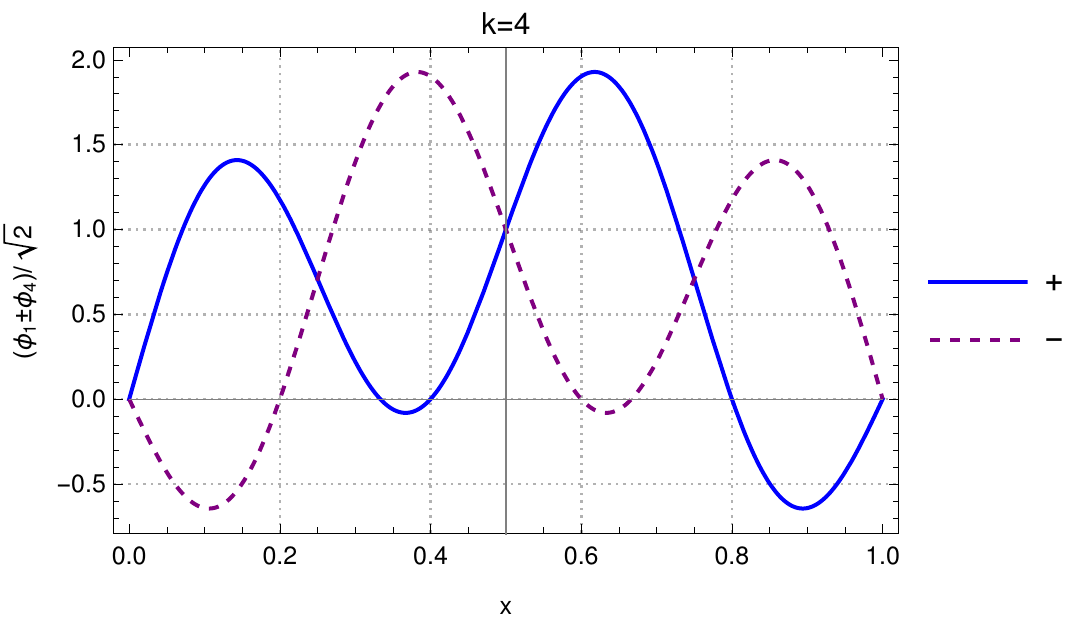}
    \caption{Localized (equivalent~\cite{Len-49}) orbitals for two independent particles in a one-dimensional box, in a state where the canonical orbitals $\phi_1$ and $\phi_k$ are occupied; for $k=2$ (top panel), $k=3$ (middle panel), and $k=4$ (bottom panel).}
    \label{fig:localized}
\end{figure}

This can be seen by orbital localization, by considering the equivalent orbitals~\cite{Len-49} $(\phi_1 \pm \phi_k)/\sqrt{2}$ instead of $\phi_1$ and $\phi_k$.
For odd $k$, the orbitals do not localize into one of the halves of the boxes as the linear combination does not change the symmetry.
For even $k$, the degree of localization decreases with increasing $k$, see Fig.~\ref{fig:localized}.

The observation that for odd $k$ there is no orthogonalization (Pauli principle) effect on the probability of having one electron in the left part of the box, and one in the right part of the box -- and there is no orbital localization in the half-boxes -- does not mean that, with a proper choice of the region, the probability to have one and only one electron cannot be made larger than for Ehrenfest urns.
For $k=3$, the probability to have one electron in the interval is 0.89 in the maximum probability domain, $x \in (0.32,0.68)$.
The price to pay for it, is that the other electron is -- with the same probability -- in spatially disjoint regions.
This quantum effect has no analogy in classical descriptions, but could be relevant, for example, in the description of time-dependent processes, such as chemical reactions.~\cite{Sav-18}


\subsection{Electron pairs: \ch{CH4}, \ch{HCCH}, \ch{F2}, \ch{XeF2}}

\begin{figure}[h]
 \includegraphics[width=0.85\textwidth]{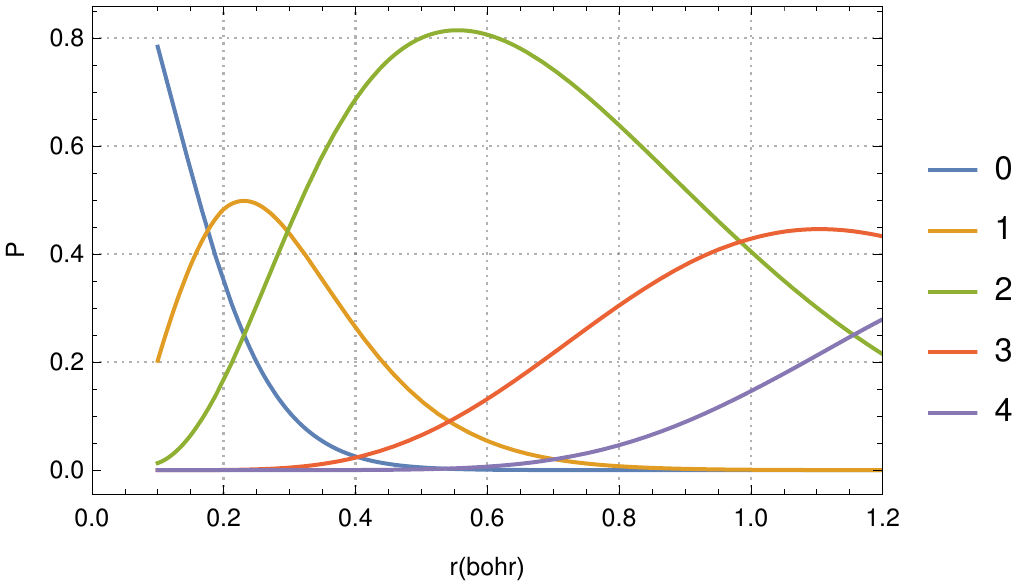}
 \caption{The probabilities of $n=0,1,\dots,4$ electrons to be present in a sphere of radius $r$ around the C nucleus in \ch{CH4}.}
 \label{fig:c-core}
\end{figure}

In order to get our bearings, let us first consider a simple example, that of the \ch{CH4} molecule.
We define the core region, by choosing $\Omega$ to be a sphere maximizing the probability of $n=2$ electrons to be inside it, see Fig.~\ref{fig:c-core}.
For a radius of 0.54~bohr, $P_{\text{core}}(n)=0.78$, while $P_{\text{core}}(n \pm 1)=0.1$.

\begin{figure}[h!]
    \centering
    \includegraphics{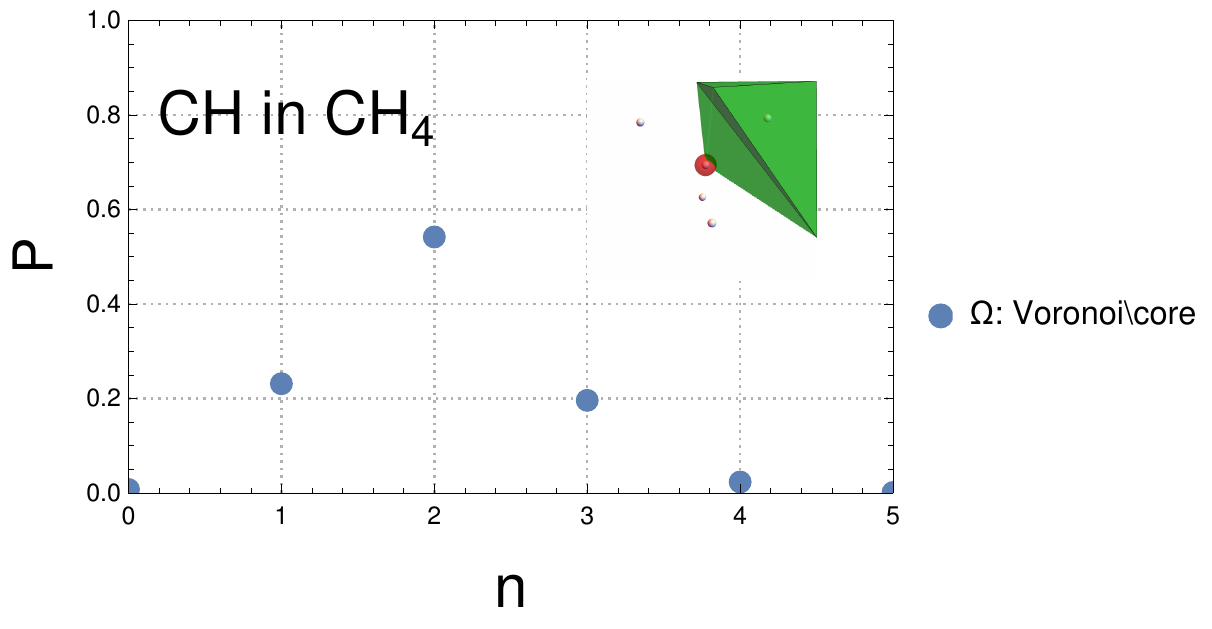}
    \caption{Probability to find $n$ electrons in the region attributed to the CH bond in \ch{CH4}. The region is shown in the top right corner of the plot, the (excluded) C core region being also shown, in red.}
    \label{fig:p-ch4}
\end{figure}

We define the valence space by excluding the sphere attributed to the core.
We divide the valence space into four Voronoi cells.
A Voronoi cell is the collection of points closest to one of the H nuclei, as shown in Fig.~\ref{fig:p-ch4}.
We associate such a region to a CH bond, and find the probability of having two electrons in it to be $P_{\ch{CH}}(n=2) = 0.55$.
This value is lower than that obtained for two electrons in the core.
At the same time, the probabilities to have one more electron in this region (or to have one less) is around $0.2$, and thus higher than for the core.
$P_{\ch{CH}}(n=2)$, it is significantly higher than given by the multiple urn model (0.31 for eight balls in four urns).
The fluctuations are thus smaller for the electronic system, than in the urns; we see it as a signature of the Pauli exclusion principle.
Nevertheless, fluctuations are present even in these (CH) bonds where the pairs of electrons are assumed to be well-localized: the probability to have two and only two electrons in the CH domain is large, but significantly different from one: the probability to have 1 or 3 electrons is smaller (0.4), but not negligible.


\begin{figure}[h!]
    \centering
    \includegraphics[height=0.29\textheight]{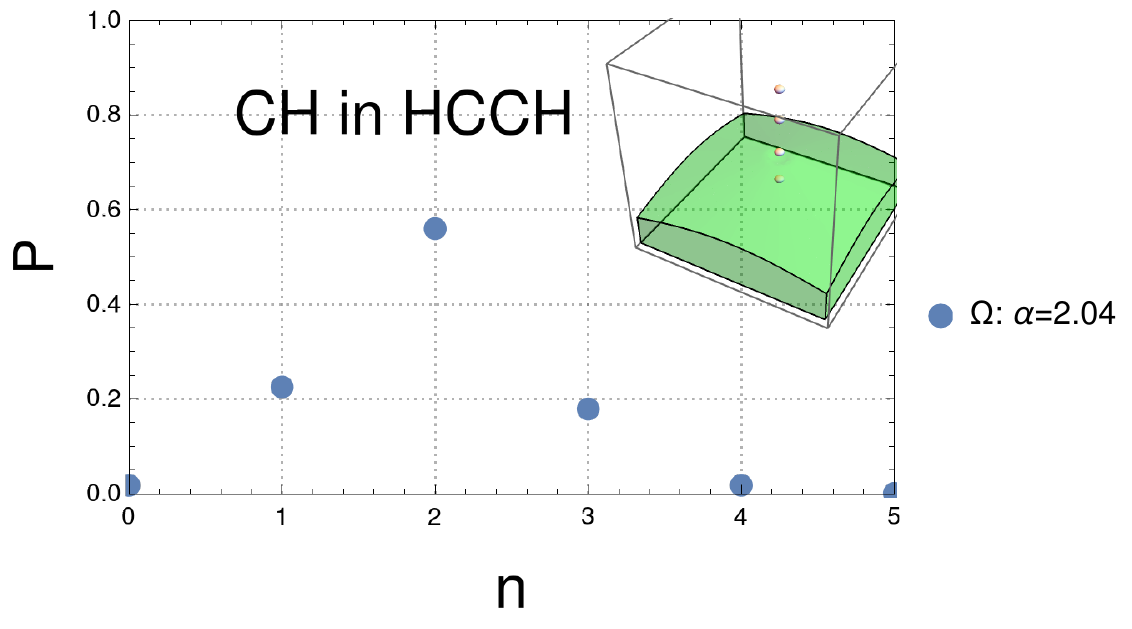} \\
    \includegraphics[height=0.29\textheight]{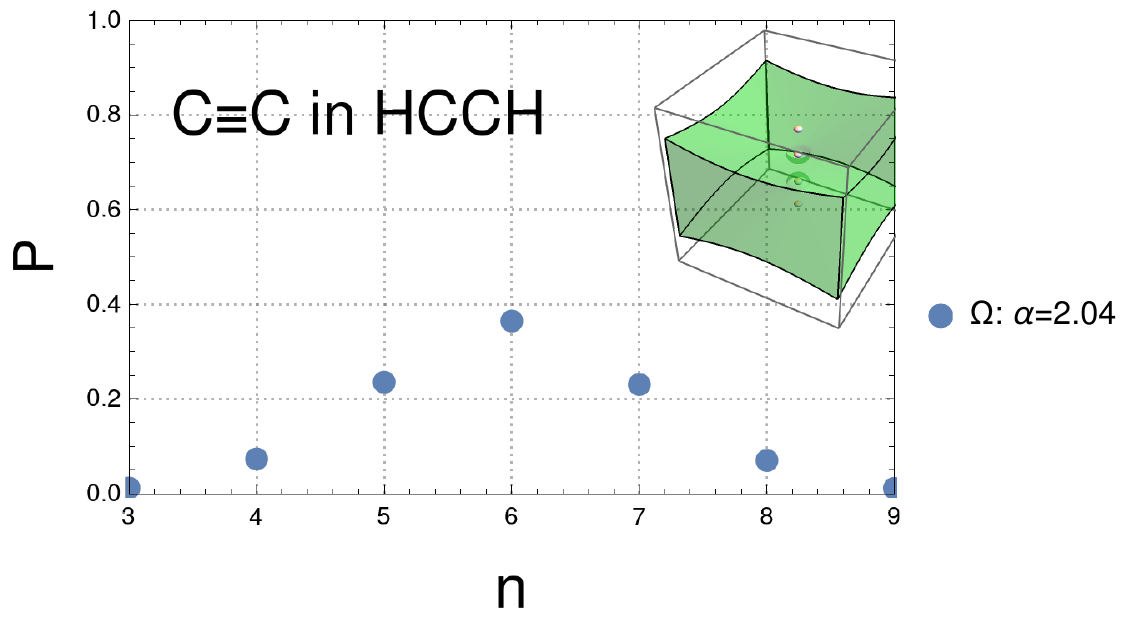} \\
    \includegraphics[height=0.29\textheight]{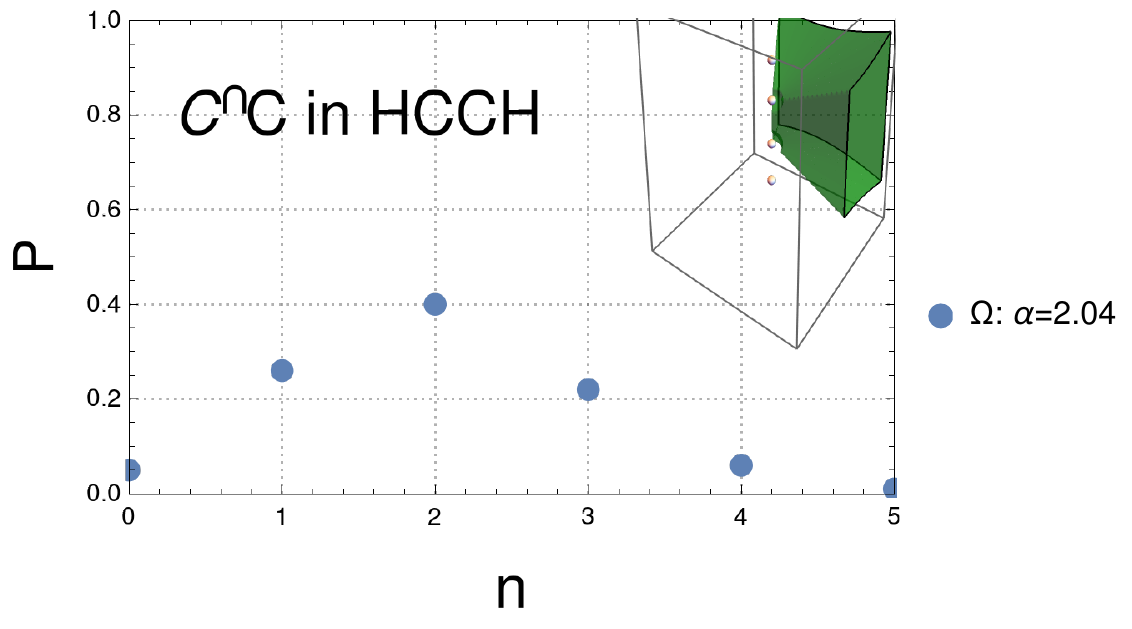}
    \caption{Probability to find $n$ electrons in the region attributed to the CH bond (top panel), the triple bond (middle panel), and the banana bond in \ch{HCCH}.}
    \label{fig:p-hcch}
\end{figure}

\begin{figure}[h!]
    \centering
    \includegraphics[height=0.29\textheight]{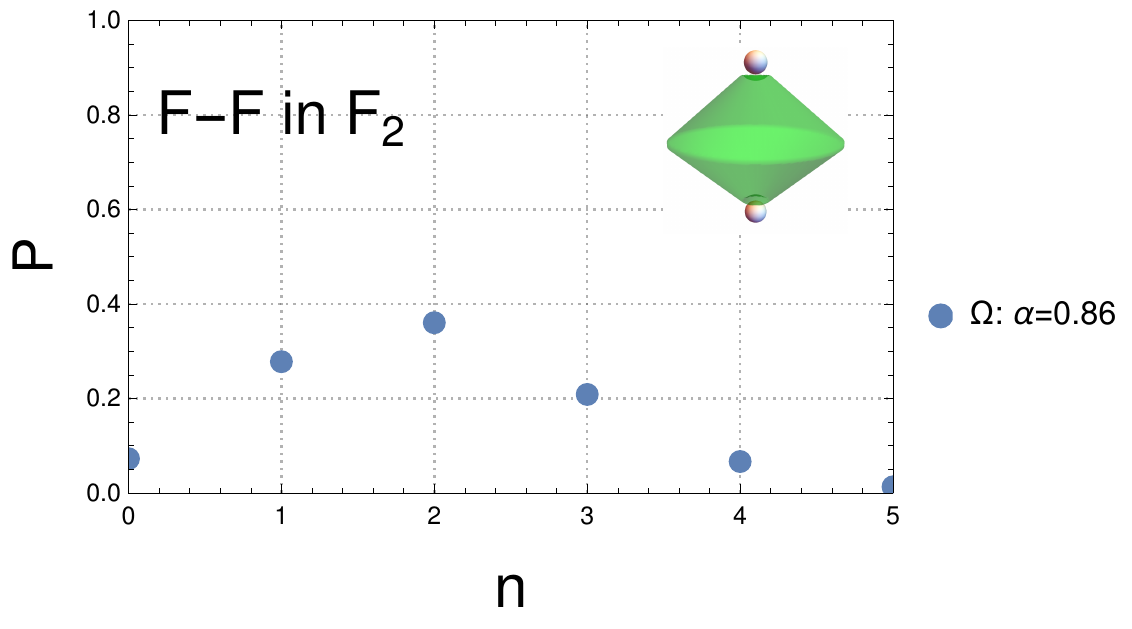} \\
    \includegraphics[height=0.29\textheight]{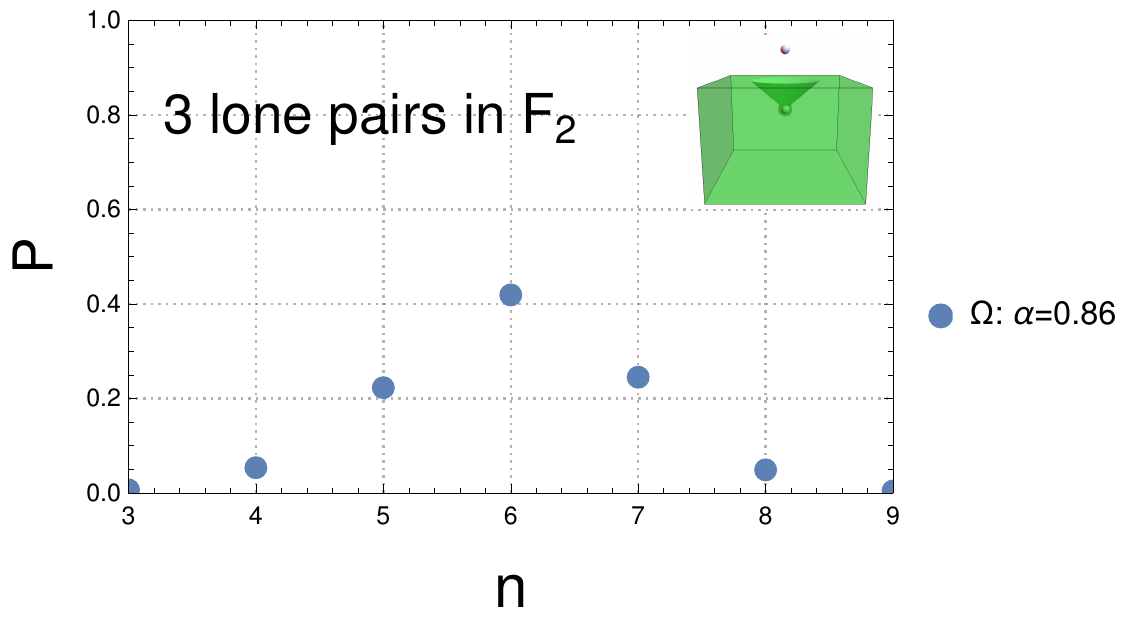} \\
    \includegraphics[height=0.29\textheight]{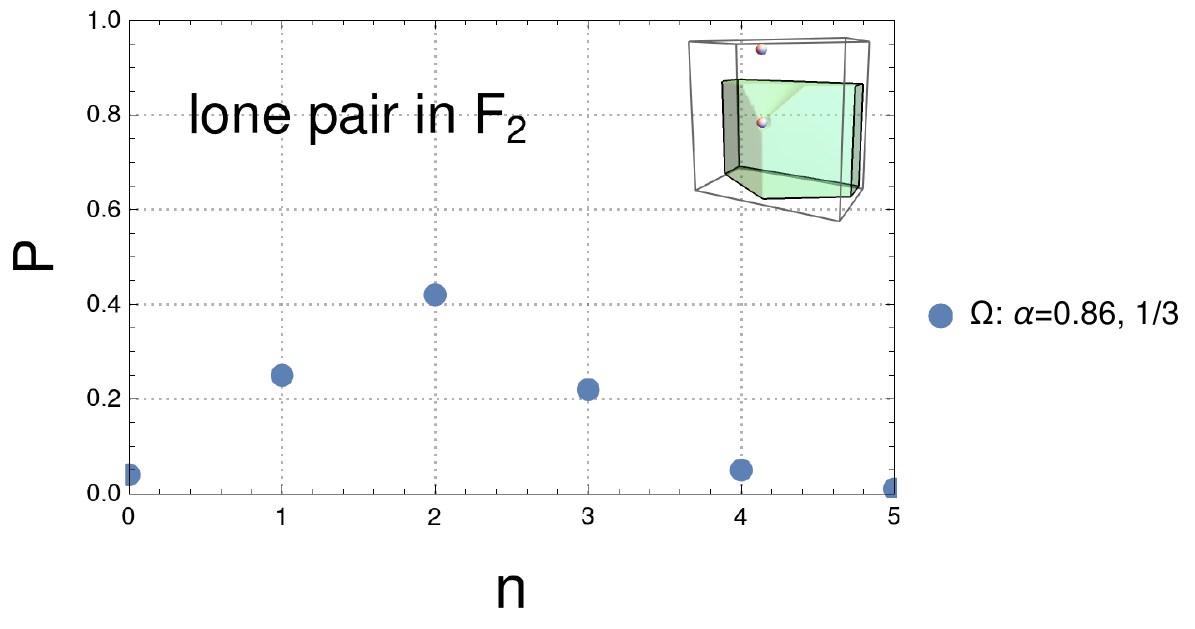}
    \caption{Probability to find $n$ electrons in the region attributed to the F-F bond (top panel), the three lone pairs on one of the atoms (middle panel), and a single lone pair in \ch{F2}.}
    \label{fig:p-f2}
\end{figure}

Let us now consider another CH bond, that in acetylene, \ch{HCCH}.
A conical region  is generated for each of the CH, while the remaining part corresponds to the CC bond (the spherical core regions being excluded, as shown in Fig.~\ref{fig:sketch}.
The probabilities obtained for the CH region are very close to those indicated for \ch{CH4}, see Fig.~\ref{fig:p-hcch}, top.

7
Other examples of electron pairs (bonds or lone pairs) are
the \ch{F - F} bond or a lone pair in \ch{F2}, see Fig.~\ref{fig:p-f2}, or one of the banana bonds in \ch{HCCH}. Fig.~\ref{fig:p-hcch}).
In these cases $P_\Omega(n=2) =0.4$ that is lower than seen for the CH bond, but still larger than in the multiple urn model.
Of course, the probability to have $n \ne 2$ in this region is increased when compared to \ch{CH4}.
$P(2 \pm 1)$ is still around 0.2.
Now, the probability to have two electrons in the region is comparable to that of having $2 \pm1 $ electrons in it.

Let us now look at the triple bond in \ch{HCCH}, and the group of three lone pairs in \ch{F2}.
After a shift of the central value (2 for one pair, 6 for three pairs), we find roughly the same distribution of probabilities as for the pairs of electrons just presented (see Figs.~\ref{fig:p-hcch} and~\ref{fig:p-f2}).
The fluctuations are significant.
The sum of the probabilities to have $n-1$ and $n+1$ electrons in the CC region is slightly higher than that of having $n=6$ electrons in it.

\subsection{Atoms: F, Xe}

\begin{figure}[h!]
    \centering
    \includegraphics[height=0.29\textheight]{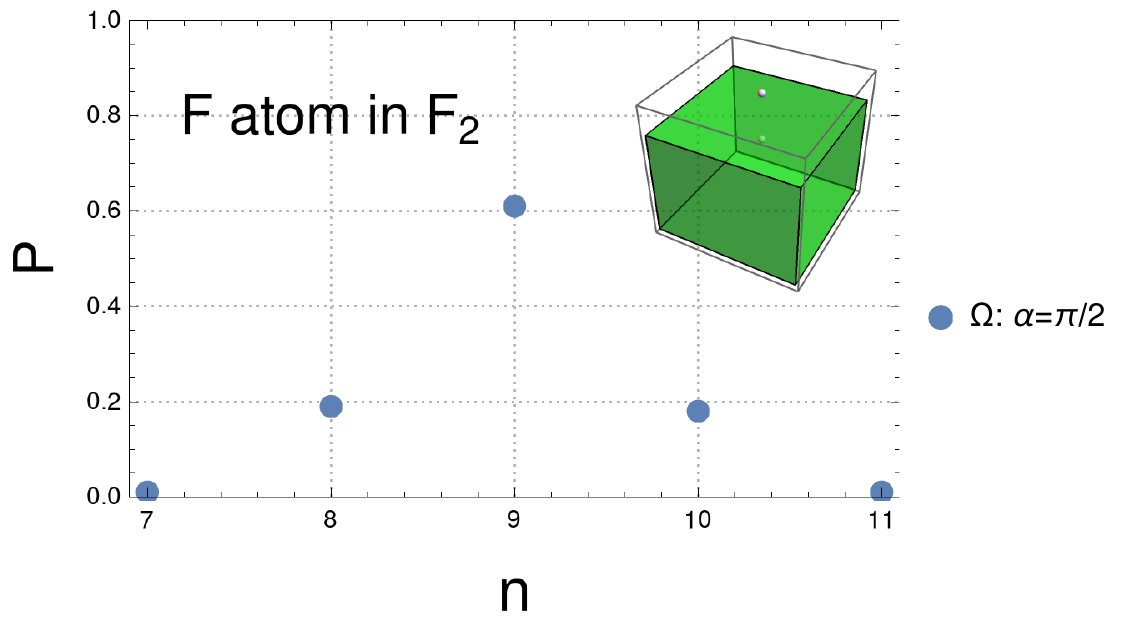}
    \caption{Probabilities for the F atom in \ch{F2}.}
    \label{fig:p-f2-f}
\end{figure}

We can define also spatial regions for atoms.
In the \ch{F2} molecule there is not much to discuss: we divide the molecule by a plane perpendicular to the line passing through the
two F nuclei, at half distance from each of these.
As we discuss atoms, let us now also include the cores.
The probability to have 9 electrons on each F atom is around 0.6, higher than in the two urn model (see Fig.~\ref{fig:p-f2-f}).


\begin{figure}[h!]
    \centering
    \includegraphics[height=0.29\textheight]{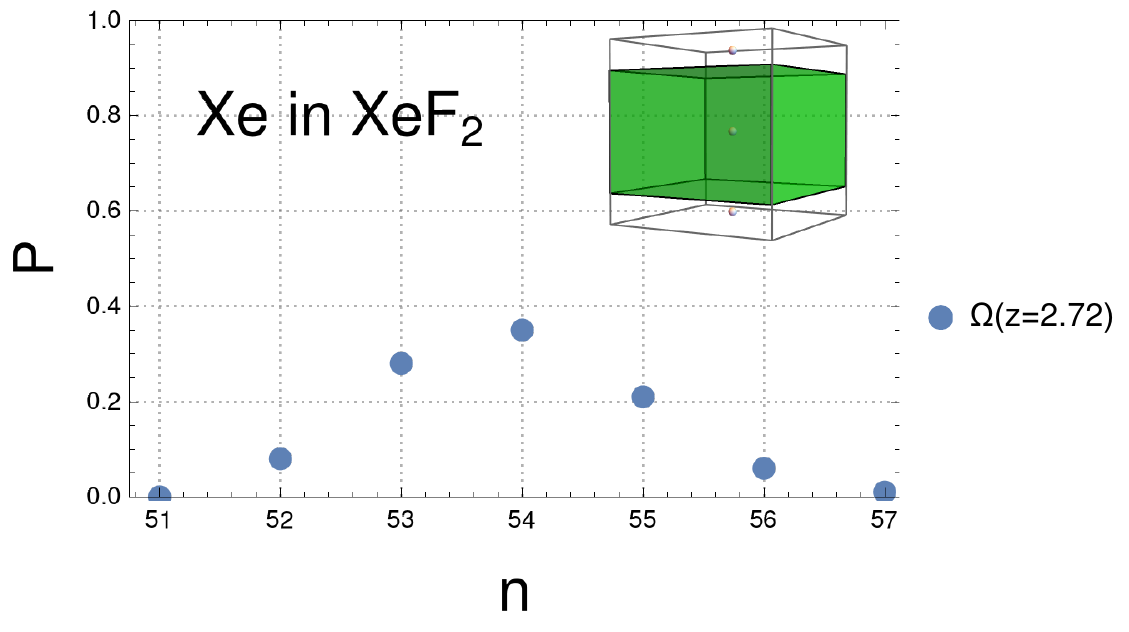} \\
    \includegraphics[height=0.29\textheight]{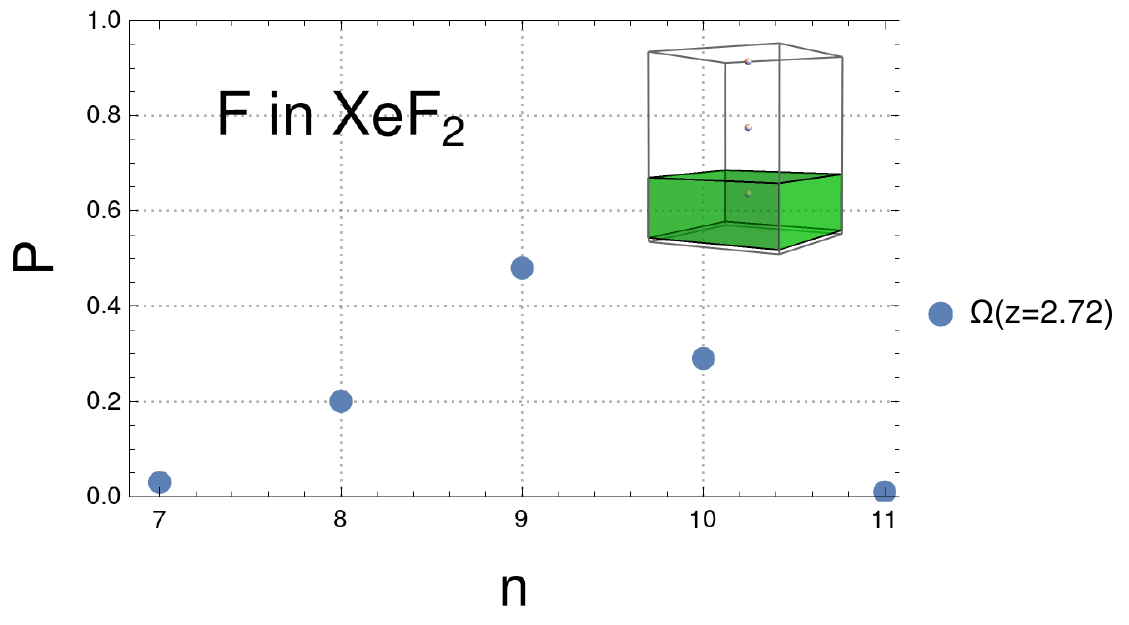}
    \caption{Probability to have $n$ electrons in the region attributed to the Xe atom (top panel), and the F atom (bottom panel) in \ch{XeF2}.}
    \label{fig:p-xef2}
\end{figure}
Let us now consider another molecule, \ch{XeF2}.
We now move two planes symmetrically around the Xe nucleus, separating the Xe atom from the F atoms.
The probability of having $n=54$ electrons within these planes is maximal ($P_{\mathrm{Xe}} = 0.35$) for $z_{\mathrm{Xe}} = 2.72$~bohr, see Fig.~\ref{fig:p-xef2}.\footnote{It is larger than the best value that is maximal for a sphere around the Xe nucleus (0.28).}
Even more pronounced than in  \ch{HCCH}, we have $P_\Omega(n) < P_\Omega(n-1)+P_\Omega(n+1)$: it is less likely to find a number of electrons equal to that expected for the Xe atom (54) than to find a different one.
Of course, we used just planes, and did not fully optimize the spatial regions to maximize $P(n=54)$, but we do not expect a qualitative change.
Further below, we argue that this might be related to the difficulty of recovering the historical chemical concepts.

\subsection{Populations}

In all cases, we find that the regions $\Omega$ maximizing the probability to have $n$ electrons in it also have a population $Q_\Omega$, Eq.~\eqref{eq:pop}, close to $n$ -- to an accuracy comparable to that we used in obtaining the probabilities.
This can be explained by the symmetry of the distributions observed.

One may wonder if it would not be better to optimize the regions to have the desired average number of electrons, and forget about fluctuations.
Finding a region that satisfies the $Q_\Omega=n$ does not seem advisable.
First, this would not detect the presence of resonant ionic structures having on average the same number of electrons as the covalent structure.
Second, there are infinitely many partitions of space that
give exactly the same average.
For example, in \ch{CH4}, we can rotate the Voronoi polyhedra that divide the space among H atoms, and still obtain the same average number of electrons in it.

The population, $Q_\Omega$, Eq.~\eqref{eq:pop}, is only the average number of electrons in a spatial domain.
We have to associate some uncertainty to it.
We could try the standard deviation, $\sigma_\Omega$, Eq.~\eqref{eq:sigma}.
When discussing Gaussian distributions, one often indicates  $2 \sigma$ as a margin of error.
One can instead state that the probability to get a result inside this confidence interval is higher than $\approx 0.95$.
The latter characterization can be used also for other probabilistic distributions.
So, let us return to our probability distributions.
From the expression of $\sigma$, Eq.~\eqref{eq:sigma}, we see that it reaches its maximum, 0.5,  only for $P_\Omega=0.5$.
So, in general, $\sigma < 0.5$, and $2 \sigma < 1$.
As the population $Q_\Omega$  is very often close to the number of electrons for which we have considered the probability, $n_\Omega$ -- that is always an integer -- there is only one integer in the interval $(Q_\Omega - 2\sigma, Q_\Omega + 2 \sigma)$.
Thus, in general, this extension of the allowed range of values does not increase the probability of having electrons beyond $P(n)$.
As we have seen, even for the clear-cut core/valence separation $P(n)$ reaches only 0.8.
So, in order to be on the safe side (of having a probability larger than 0.95), to accept in our picture at least the transfer of at least one electron between spatial regions.

\subsection{Joint probabilities}

We can extend the previous discussion to multiple regions.
Let us come back to the \ch{CH4} molecule.
The numbers given for \ch{CH4} correspond to the probability to have two electrons in one CH bond region, no matter how the electrons are distributed in the other regions.
The joint probability to have two electrons in each of the CH regions is lower (0.20).
This distribution, $2|2|2|2$ is the most probable distribution of electrons over the four regions.
However, exchange of one electron between two regions can occur in many ways ($1|3|2|2, 3|1|2|2, 1|2|3|2, \dots$).
Summing up all the probabilities related to all these distribution yields a value more than twice as large (0.44).
However, the value obtained for $2|2|2|2$ is five times larger than the probability of obtaining two balls in each of four urns.

This low joint probability of the most probable distribution of electrons is no exception.
For example, in \ch{HCCH}, the partition $2|6|2$ is the most probable ($P=0.2$).
However, the partitions where one electron is transferred between the two regions (like $1|7|2, 3|5|2$) have a probability of 0.3.
For the \ch{F2} molecule, the valence partition $6|2|6$ yields a joint probability of only 0.1, while those with one electron transferred (like $7|1|6$, $5|3|6$, or $5|2|7$) produce 0.3.
In this case the transfer between lone pair regions on different atoms that is more important here than, e.g., between the CH regions in \ch{HCCH}: the lone pair regions have a common border in our partition.
In fact the probability of $5|2|7$ and that of $7|2|5$ is smaller than that of $6|2|6$, but also around 0.1.

The probability to have two electrons in each of the valence electron pairs (for example, two in each of the CH bond regions, and two in each of the banana bond regions in \ch{HCCH}) would be even lower.
This lowers the probability of having a Lewis-type structure.
Furthermore, were we to take into account the probability of electrons to move between the core and valence region, the probabilities would further drop.
For example, that of having one of the six electrons of the region attributed to the CC triple bond being in one of the two C core regions, $2|5|2$, is 0.06.

\subsection{Conditional probabilities}

One way to understand the low values for joint probabilities is by starting to assume that the probability to have $n_A$ electrons in $A$ and $n_B$ in $B$ are independent of each other,
\begin{equation}
    0 < P_A(n_A) P_B(n_B) < P_A(n_A) \; \mathrm{or} \; P_B(n_B) < 1.
\end{equation}
However, in general, independence cannot be assumed.
For example, when an electro\-ne\-ga\-tive element, say, F takes a charge, it takes it from somewhere.
We have a probability that is was being taken from some other spatial domain.
So we ask: What is the probability to have $n_A$ in region $A$ (say, 54-1 electrons in the Xe region of \ch{XeF2}), given that there are $n_B$ in region $B$ (say, 9 in one of the F regions)?
If the fluctuations were due to independent events, it would be easy to answer, because $P_{A|B}=P_A$.

\begin{figure}[h!]
    \centering
    \includegraphics[height=0.29\textheight]{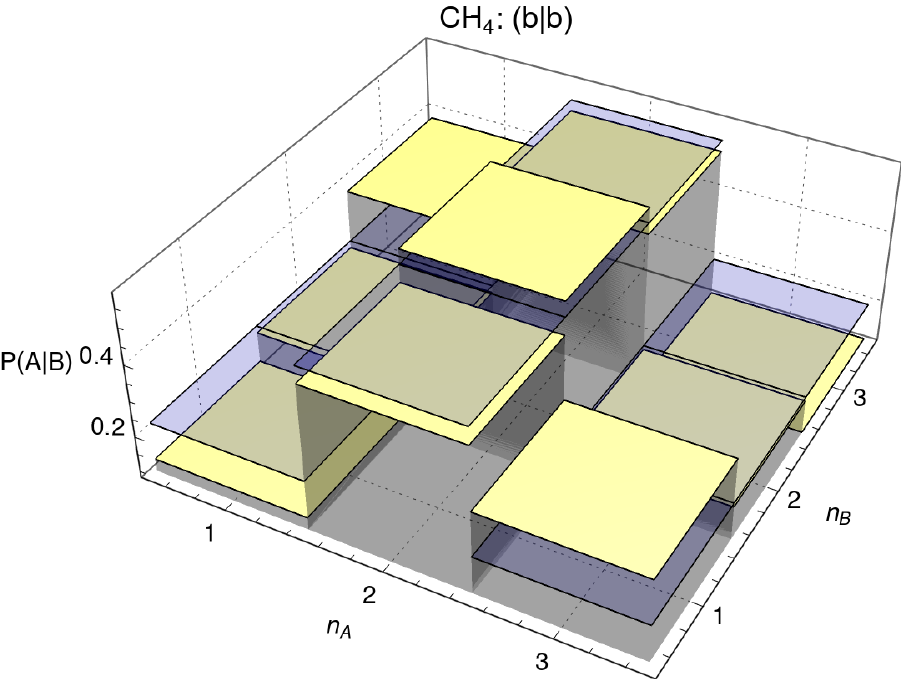} \\
    \caption{The conditional probability $P_{A|B}$ of having $n_A$ electrons in region $A$ given that there are $n_B$ electrons in region $B$, for \ch{CH4}; $A$ and $B$ correspond to two different CH bond regions. The horizontal semi-transparent blue planes show $P_A(n_A)$. }
    \label{fig:pcond-ch4}
\end{figure}

\begin{figure}[h!]
    \centering
    \includegraphics[height=0.27\textheight]{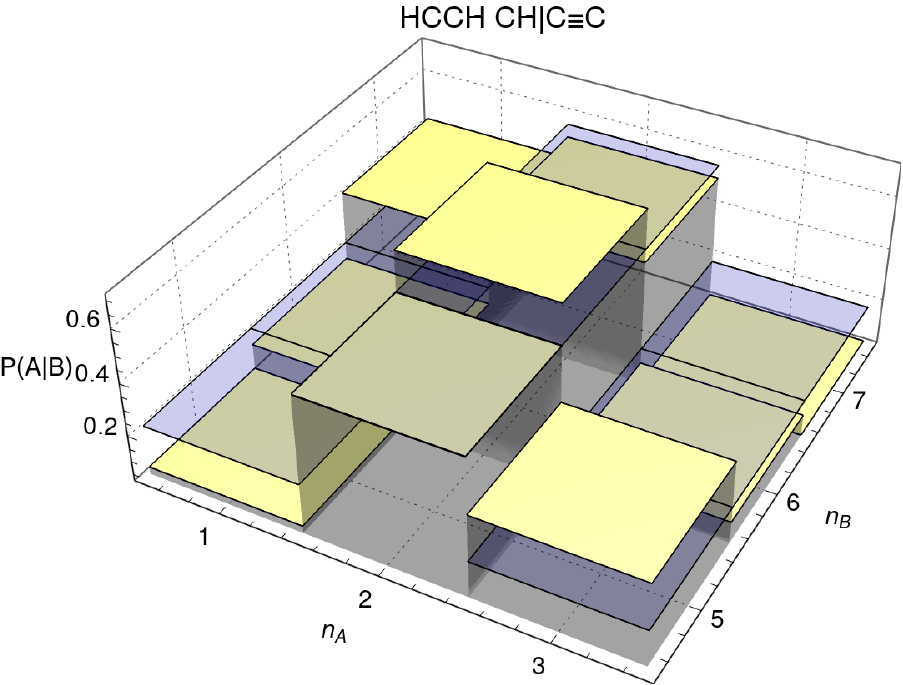} \\
    \includegraphics[height=0.27\textheight]{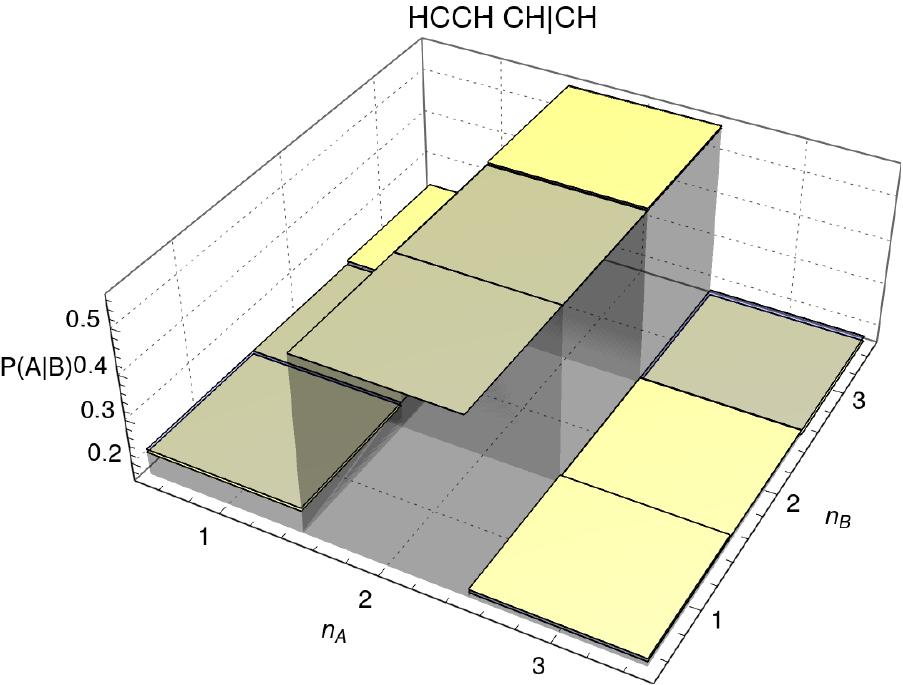} \\
    \includegraphics[height=0.27\textheight]{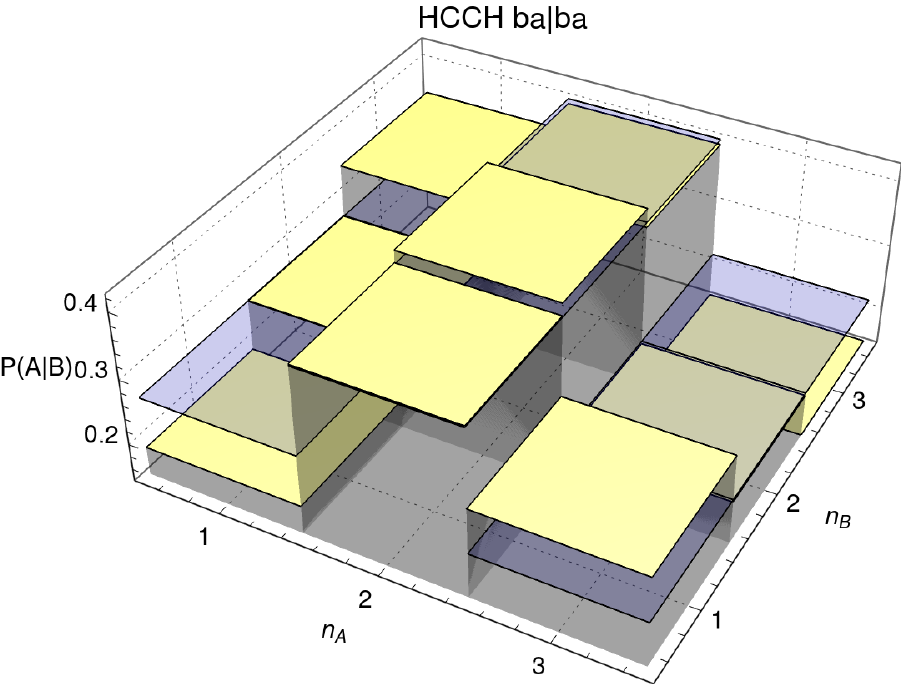}
    \caption{The conditional probability $P_{A|B}$ of having $n_A$ electrons in region $A$ given that there are $n_B$ electrons in region $B$, for \ch{HCCH} (top: $A$=CH, $B$=CC, middle $A,B$ are two different CH regions, bottom: $A,B$ are two different banana bonds. The horizontal semi-transparent blue planes show $P_A(n_A)$. }
    \label{fig:pcond-hcch}
\end{figure}

\begin{figure}[h!]
    \centering
    \includegraphics[height=0.29\textheight]{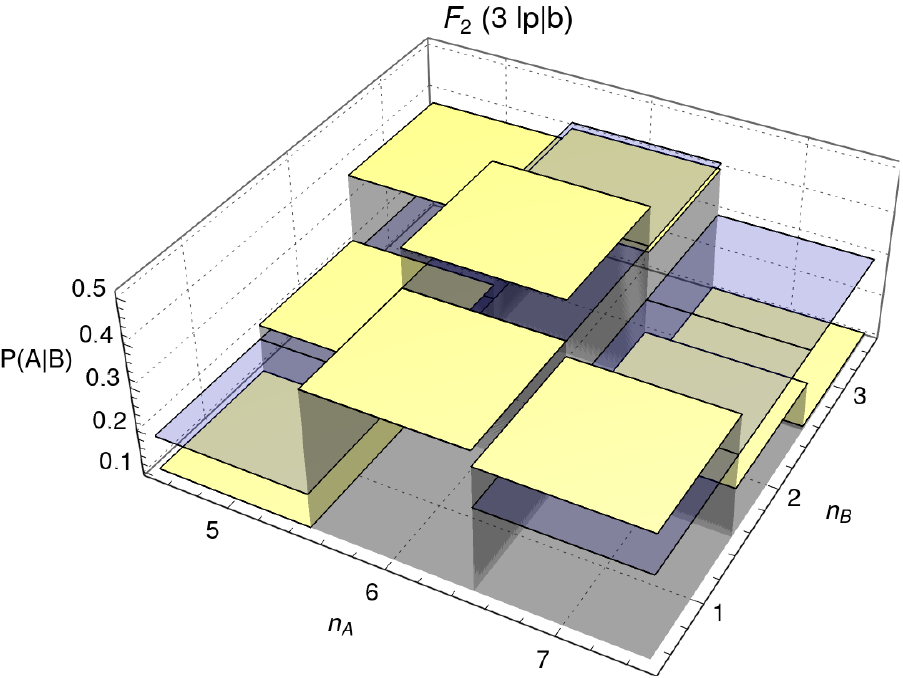} \\
    \includegraphics[height=0.29\textheight]{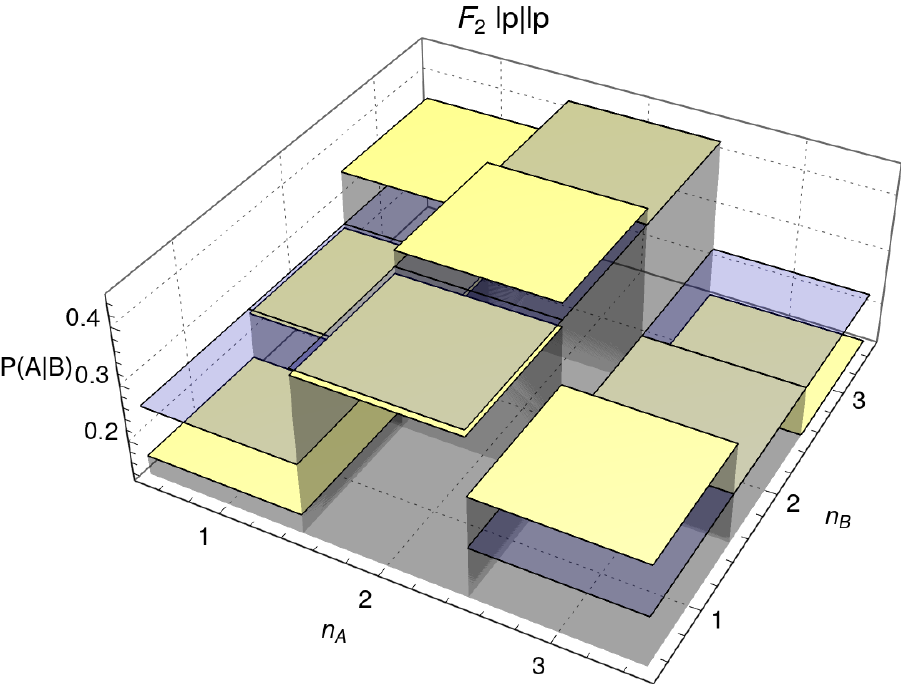} \\
    \caption{The conditional probability $P_{A|B}$ of having $n_A$ electrons in region $A$ given that there are $n_B$ electrons in region $B$, for \ch{F2} (top: $A$ is the group of 3 lone pairs, and $B$ is the bond,  bottom: $A,B$ are two different lone pairs on the same atom. The horizontal semi-transparent blue planes show $P_A(n_A)$. }
    \label{fig:pcond-f2}
\end{figure}

Let us start with the simple case when regions $A$ and $B$ correspond to CH bond regions in \ch{CH4}, see Fig.~\ref{fig:pcond-ch4}.
Let us consider $n_A=2$ and $n_B=2$, the chemically assumed values.
The probability to have the chemically assumed value in $B$ increases that of having the chemically assumed value in $A$ (compared to independent events).
Furthermore, having $n_B+1$ also increases the probability to have $n_A-1$ in the neighbouring region.
Evidently, this holds for $n_B-1$ and $n_A+1$.
However, a simultaneous increase or decrease of the number of electrons in neighbouring regions is less favorable than for independent particles. $P_{A|B}(n_A + 1, n_B +1) < P_A(n_A+1)$ and $P_{A|B}(n_A - 1, n_B - 1) < P_A(n_A - 1)$.
A similar pattern is seen, e.g., in \ch{HCCH} when $A$ is the CH region, and $B$ the CC region (see Fig.~\ref{fig:pcond-hcch}, top panel), or in \ch{F2} when when $A$ is the region of the three lone pairs, and $B$ that of the bonding region.
We suspect these trends to be quite common.

A further pattern is present when looking at distant pairs, like the two CH bonds in \ch{HCCH} (Fig.~\ref{fig:pcond-hcch}, middle panel).
We see practically no reciprocal influence.

An intermediate pattern shows up, when looking how one banana bond influences the other (Fig.~\ref{fig:pcond-hcch}, lower panel), or how lone pairs on the same atom interact (in \ch{F2}, Fig.~\ref{fig:pcond-f2}, lower panel): $P_{A|B}(n_A=2,n_B=2) \approx P_A(n_A=2)$.
However,  $P_{A|B}(n_A=1,n_B=3) > P_A(n_A=1)$ and $P_{A|B}(n_A=3,n_B=1) > P_A(n_A=3)$.
This increased electron transfer between two electron regions only reminds of ionic structures, but it is not: it happens on the same atom (in \ch{F2}), or within the triple bond (in \ch{HCCH}).

\begin{figure}[h]
    \centering
    \includegraphics{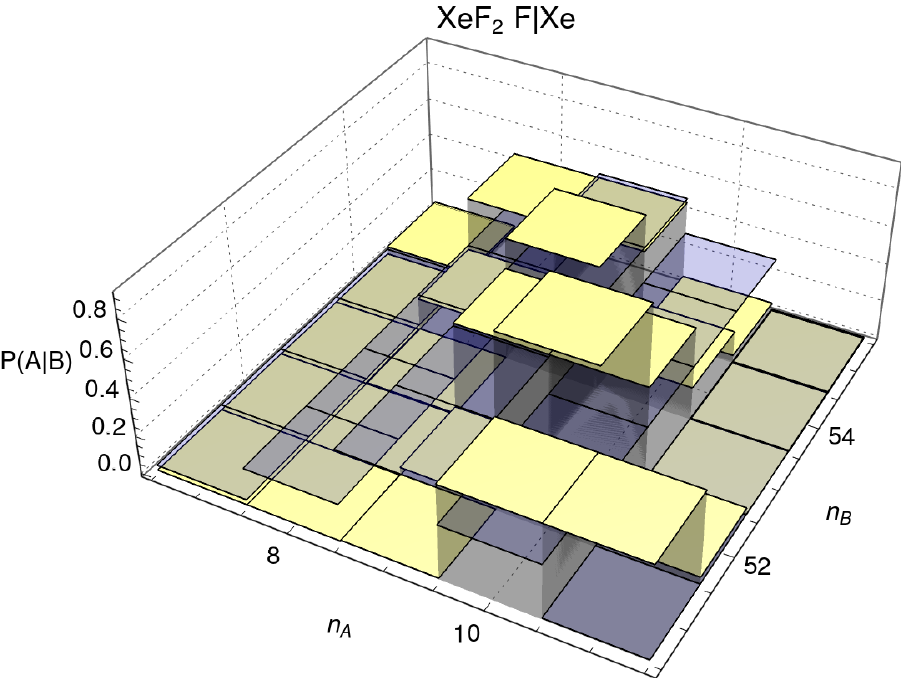} \\
    \includegraphics{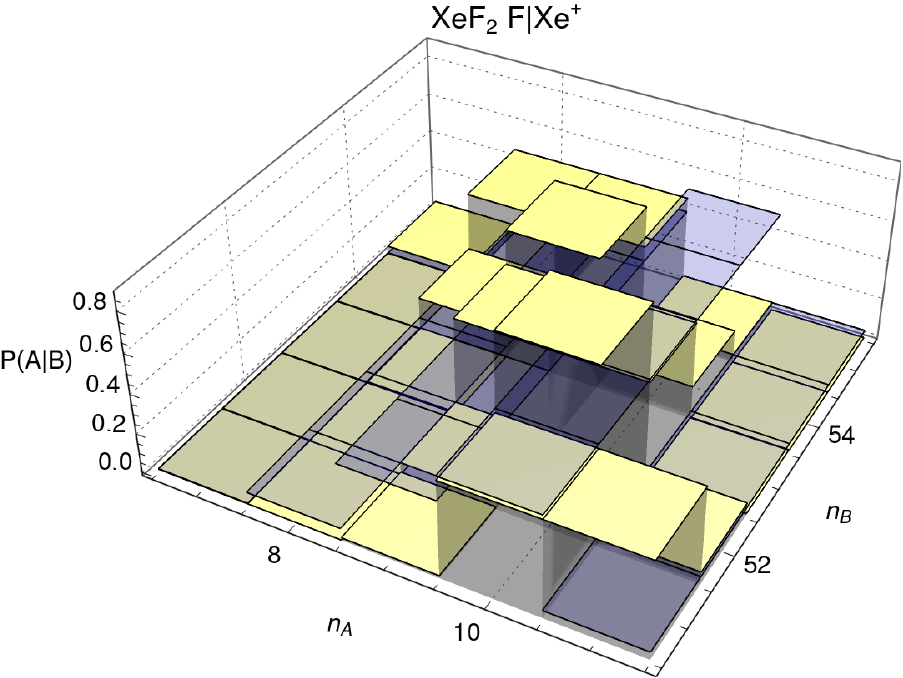}
    \caption{ Conditional probability in \ch{XeF2}, for finding $n_A=7, 8,\dots,11$ electrons in the F atom regions, given that there are $n_B=51,52,\dots,55$ electrons in the Xe atom region (top panel), or \ch{Xe+} ion region (bottom panel).}
    \label{fig:pcond-xef2}
\end{figure}

Finally, let us look at \ch{XeF2} (see Fig.~\ref{fig:pcond-xef2}, upper panel).
We consider the atomic regions, Xe and F and notice also a significant increase in cases that do not fit  at all in the ``standard'' cases discussed above.
For example, for $n_B-1=53$ we have an increase on $n_A+1$, but also on $n_A$.
Furthermore, we notice an significant increase for $n_B-2$ on $n_A+1$.

As the transfer of electrons occurs between atoms, it is more easy to describe it by ``ionic structures'' that get more weight than what one would expect by simple ``urn-like'' fluctuation.
This leads us to ask whether it would be not convenient to use ions as a reference, and not atoms.

\section{Choosing the significant region}

\subsection{Alternatives}

\begin{figure}[h]
    \centering
    \includegraphics{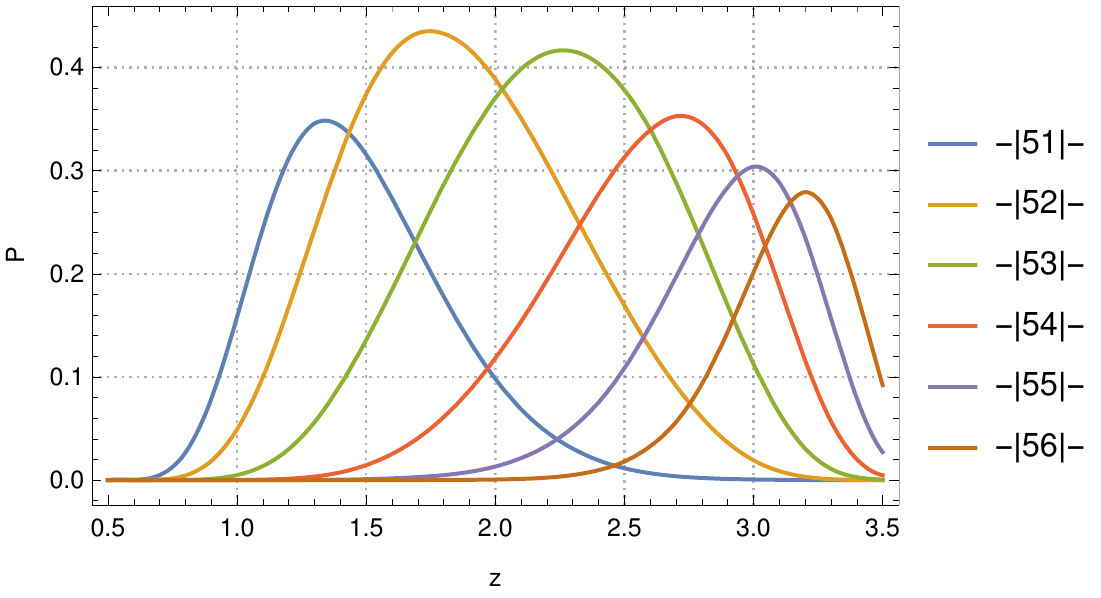}
    \includegraphics{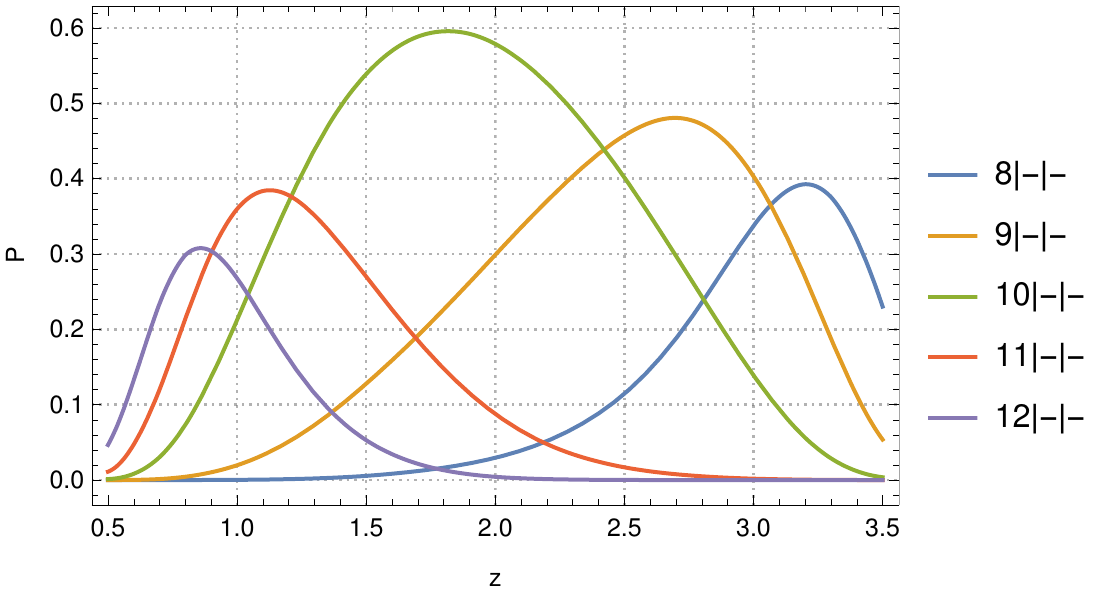}
    \caption{Probability to have $n$ electrons between two planes at $\pm z$, in \ch{XeF2}. The Xe nucleus is at $z=0$, the $F$ nucleus at $z\approx 3.74$. The curves marked $-|n|-$ (top panel) show the probability to have $n$ electrons in the region containing the Xe nucleus, the curves marked $n|-|-$ (bottom panel) show the probability to have $n$ electrons in a region containing one of the F nuclei.
    }
    \label{fig:probz-xef2}
\end{figure}

Fig.~\ref{fig:probz-xef2} shows how probabilities evolve when changing the positions of the planes perpendicular to the $z$-axis, and symmetrically placed with respect to the origin (the position of the Xe nucleus, the F nuclei being at $z= \pm 3.74$~bohr).
We see that there is indeed a maximum in the probability of having $n=54$ electrons in a region around the Xe nucleus, but the probability to have $n=53$ and even $n=52$ is larger.
The effect is even more pronounced if we look at the region around one of the F nuclei. The probability to have $n=10$ is clearly more pronounced than that of having $n=9$.

\begin{figure}[h]
    \centering
    \includegraphics{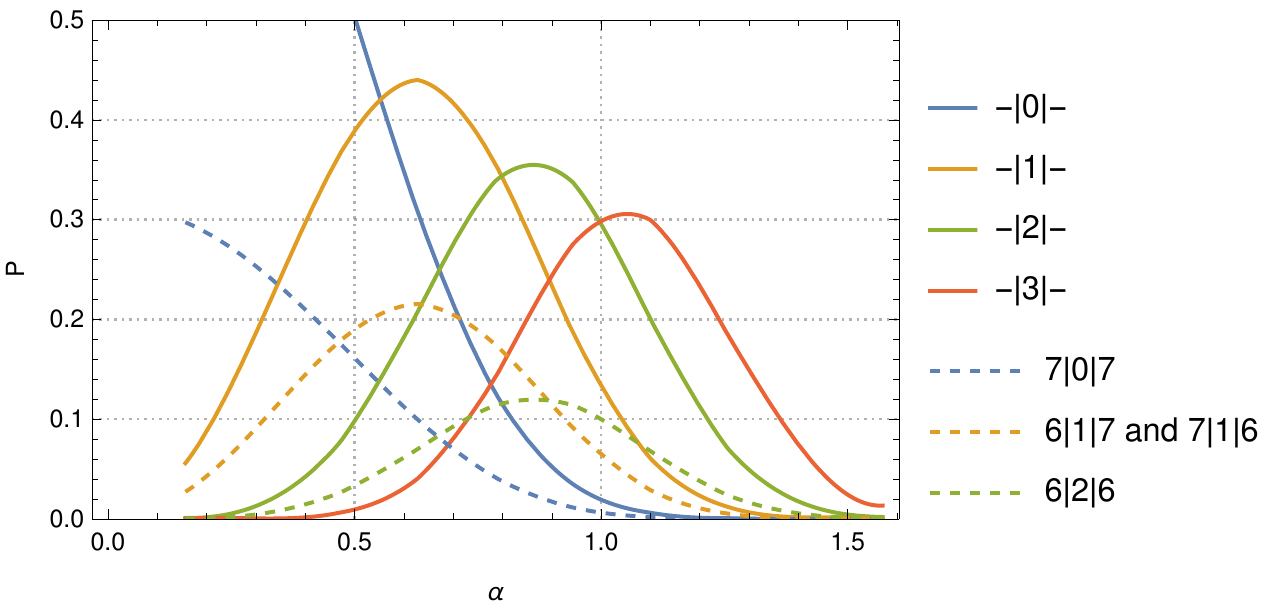} \\
    \includegraphics{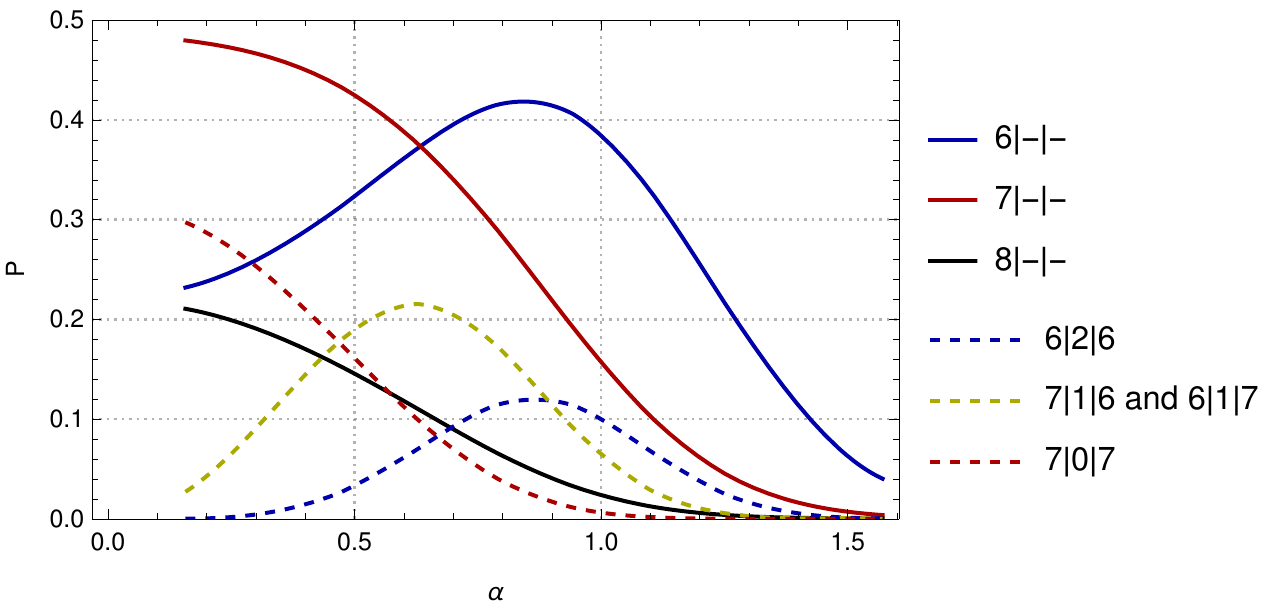}
    \caption{\ch{F2} molecule, probability for the regions corresponding to the FF bond (top), and the three lone pairs (bottom); full lines are for having 1, 2, or 3 electrons in the bond region, or 6, 7, 8 for the region of three lone pairs. The dashed lines correspond to specific partitions.}
    \label{fig:probz-f2}.
\end{figure}

A similar feature is present when looking at the \ch{F2} molecule, Fig.~\ref{fig:probz-f2}.
The probability to have one electron in the region described by cones between the two nuclei can be made larger than that of having two.
Correspondingly, the probability of having seven electrons in the region attributed to three lone pairs on an atom is also getting larger than the best we can get for having six.~\footnote{The probability to have no electrons in a region of vanishingly small volume can be also seen in the figure: it is -- trivially -- equal to 1.}
This is in contrast to what we have seen for the core-valence separation (shown for \ch{CH4}, Fig.~\ref{fig:c-core}) where the expected values were also the \alert{optimal} ones.

\begin{figure}[h]
    \centering
    \includegraphics{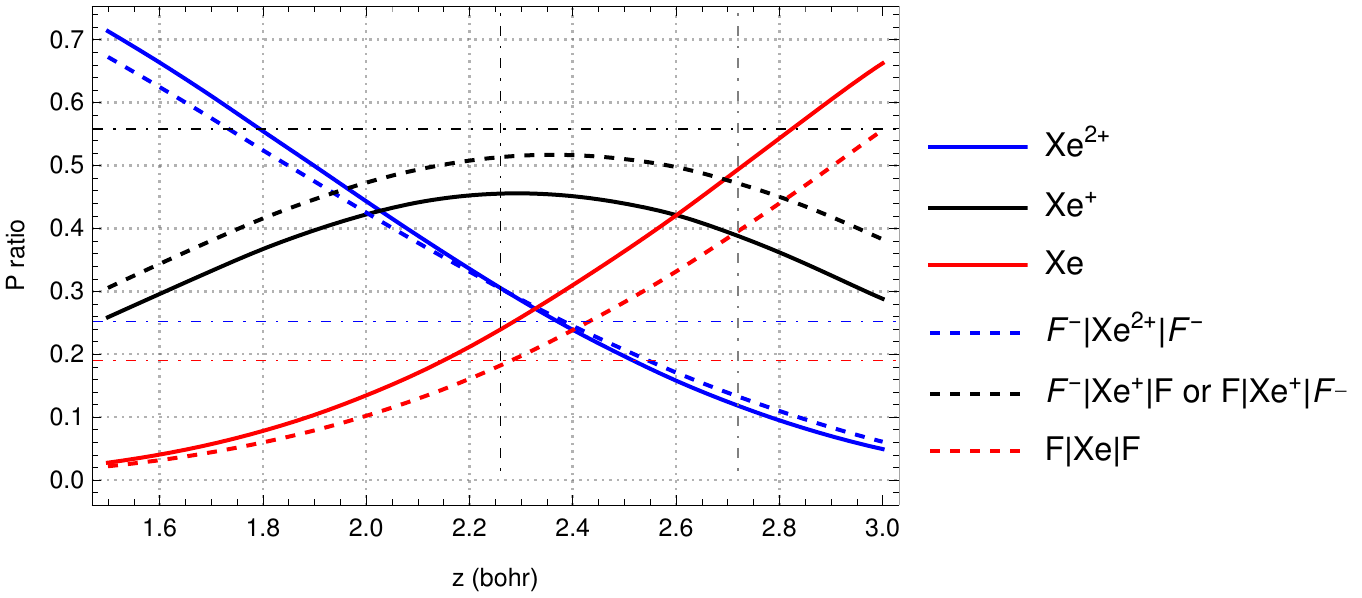}
    \caption{ Ratio of probabilities as changing with the position of the planes defining the regions in the \ch{XeF2} molecule; for $\Omega$ between $-z$ and $z$, around the Xe atom. $P_\Omega(n)/\sum_i P_\Omega(n_i)$, $n_i=$ 52 (blue), 53 (green), 54 (red), full curves. The dashed curves, show the relative contributions of the probabilities for the  partitions 10|52|10 (blue), 9|53|10 together with 10|53|9 (green), and 9|54|9 (red). The vertical dot-dashed lines indicate the values of $z$ for which $P(n=53)$ and $P(n=54)$ are maximal. The horizontal dot-dashed lines show weights obtained in valence bond calculations.~\cite{BraHib-13} }
    \label{fig:vb-xef2}
\end{figure}

As we have the signature of ionic structures, let us compare the evolution of  the probabilities of having different numbers of electrons in a region, as we change the region is changed.
\alert{From our data we find that we can have more ways to distribute electrons between regions than one would normally assume in valence bond calculations, that try to keep to minimal, significant distributions.
In order to make comparisons with valence bond calculations let us restrict ourselves to a given set of integers, showing up in some reference valence bond calculation,  $\mathcal{I}$ and consider
}
\begin{equation}
    w_\Omega(n)=\frac{P_\Omega(n)}{\sum_{k \in \mathcal{I}} P_\Omega(k)}, \; n \in \mathcal{I}
\end{equation}

Of course, $w_\Omega$ changes with $\Omega$.
Let us consider \ch{XeF2}, and displace the planes that separate the region around the Xe nucleus (placed at the origin) from the F nuclei (placed at $z=\pm 3.74$~bohr), see Fig.~\ref{fig:vb-xef2}.
We consider $\mathcal{I}=\{52,53,54\}$.
Let us choose, to start with a region between $-z$ and $z$ chosen to maximize $P(n=52)$.
As the planes get further away from Xe nucleus, the weight of $P(n=52)$ decreases, to let $P(n=53)$ increase.
As $z$ further increases, $P(n=53)$ arrives to a maximum.
Continuing to increase $z$, the contribution of $P(n=53)$ declines and lets $P(n=54)$ increase.
We can look, for the same process, at the probabilities for the region around a F atom ($\mathcal{I}={9,10}$).

Instead of looking at the ratios $w$ for $P(n)$, we can look at the ratio of the joint probabilities corresponding to partitions, e.g., $n_1 | n_\mathrm{Xe} | n_2$ that of having $n_1$ electrons in one F region, $n_\mathrm{Xe}$ in the region around Xe, and $n_2$ around the other F nucleus.
We see (in Fig.~\ref{fig:vb-xef2}) that the two types of ratios do not differ much in spite of $P_{A \cap B} $ being lower than $P(A)$.

\begin{figure}[h]
    \centering
    \includegraphics{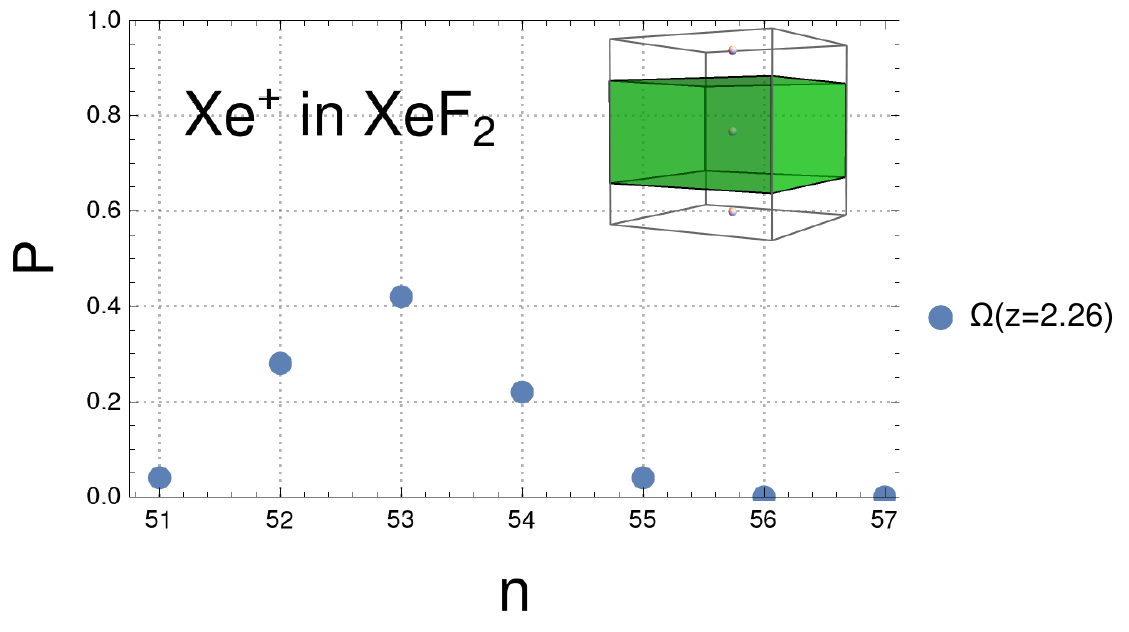} \\
    \includegraphics{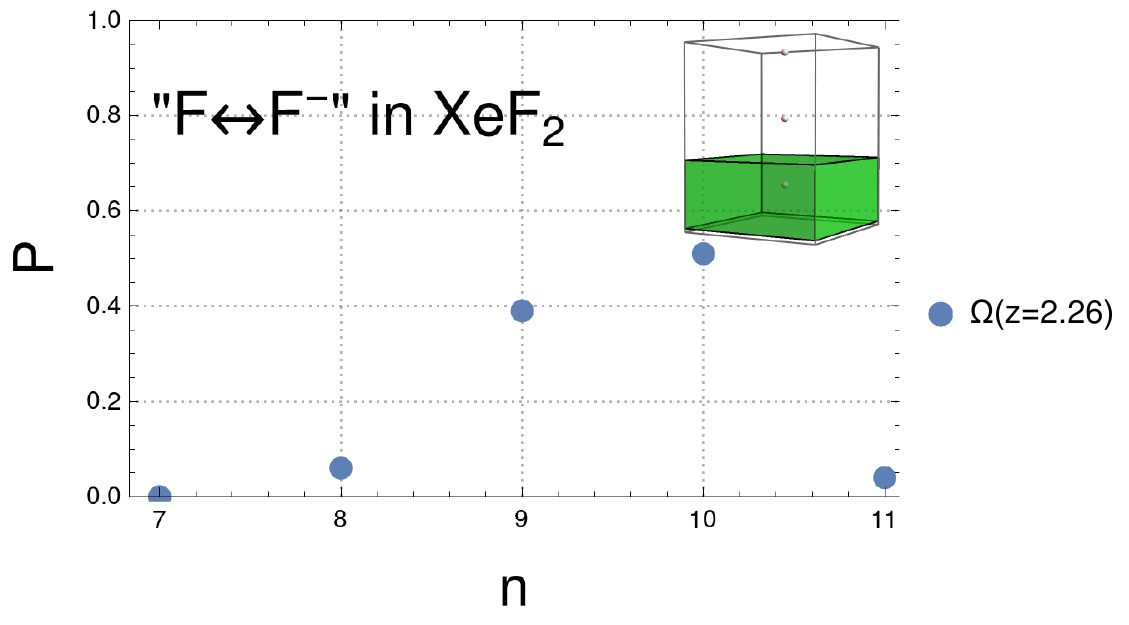}
    \caption{Probability to find $n$ electrons in the region attributed to the \ch{Xe+} ion (top panel), and to \ch{F-} $\leftrightarrow$ \ch{F}  (bottom panel) in \ch{XeF2}.}
    \label{fig:p-xepf2}
\end{figure}

We find, that for a certain value of $z$ our $49,406$ determinant wave function produces ratios of probabilities ($w_\Omega$) that are pretty close to the weights obtained in valence bond calculations.~\cite{BraHib-13}
The value of $z$ where this happens is slightly larger than that that we would have attributed to \ch{Xe+}.
This indicates that what we call a Xe atom, is, in fact, closer to a \ch{Xe+} ion.
So let us better consider the probabilities (see Fig.~\ref{fig:p-xepf2}) for the domain that maximizes the probability of having 53 electrons around the Xe nucleus (as in Fig.~\ref{fig:probz-xef2}).
At the same time, this means that we should consider having in the region around the F nucleus either 9 or 10 electrons,
F~$\leftrightarrow$~\ch{F-}.

\subsection{Effective number of parties}

We are now confronted with the existence of alternative descriptions.
In fact, already Lewis saw the need for it, and used the term tautomerism.~\cite{Lew-16}
Choosing one of the spatial region is in detriment for another.
For the example above, the optimal region \alert{for} Xe is not the best for \ch{Xe+}.
The $P(53)$ and $P(54)$ are different in the two regions.
There is an arbitrariness in the choice of the region: \alert{One person may find it convenient to choose the \ch{Xe} one, another person the \ch{Xe+} one.}
Of course, we can optimize regions by weighting their importance.
The choice of these weights has also a certain degree of arbitrariness.
In some cases, chemical intuition can decide about having equal weights, e.g., F and \ch{F-}, as
equivalent to the choice of \ch{Xe+}, see Fig.~\ref{fig:p-xepf2}.

Another possibility is to choose the region that minimizes the number of alternative structures.
In the study of political systems, the concept of ``effective number of parties'' exists: if a party is weakly represented in a parliament, it weighs less.
Laakso and Taagepera~\cite{LaaTaa-79} use for it
\begin{equation}
    \mathcal{N}=\frac{1}{\sum_n \mathcal{P}_n^2}
\end{equation}
where $\mathcal{P}_n$ is the fractional number of seats occupied by a party, and $N$ is the total number of parties present in the parliament.
$\mathcal{N} \rightarrow 1$ if one party dominates, and $\mathcal{N} =N$ if all parties have an equal number of seats.
Of course, this definition can be used for other types, by replacing $\mathcal{P}_n$ by probabilities.
Let us try to use this definition in the case of \ch{XeF2}.
\begin{enumerate}
    \item Let us use the probabilities of $n_1|n_{Xe}|n_2$ (to have $n_{Xe}$ electrons in the Xe atom region, $n_1$ in that of one of the F atoms, and $n_2$ in that of the other F atom).
If we use the partition to maximize to probability to have a Xe atom, we get $\mathcal{N}=8$, dominated by  F|Xe|F, \ch{F-}|\ch{Xe+}|\ch{F} and \ch{F}|\ch{Xe+}|\ch{F-}, \ch{F+}|\ch{Xe-}|\ch{F} and \ch{F}|\ch{Xe-}|\ch{F+}, and \ch{F-}|\ch{Xe^{2+}}|\ch{F-}, \ch{F-}|\ch{Xe}|\ch{F+} and \ch{F+}|\ch{Xe}|\ch{F-}.
If we use the partition to maximize to probability to have a \ch{Xe+} ion, we get $\mathcal{N}=6$, dominated by \ch{F-}|\ch{Xe^{2+}}|\ch{F-},  \ch{F-}|\ch{Xe+}|\ch{F} and \ch{F}|\ch{Xe+}|\ch{F-}, F|Xe|F, \ch{F-}|\ch{Xe}|\ch{F+} and \ch{F+}|\ch{Xe}|\ch{F-}.
This eliminated the partitions requiring \ch{Xe-}: and thus reduced the number of descriptors needed.
     \item It may be more reasonable to group together the chemically equivalent descriptions, e.g., \ch{F-}|\ch{Xe+}|\ch{F} and \ch{F}|\ch{Xe+}|\ch{F-}.
This way, $\mathcal{N}$ is reduced, because we count such pairs only once.
Also, the importance is modified, for example, for $n_1 \ne n_2$ instead of having $P(n_1, n_{Xe}, n_2) = P(n_2, n_{Xe}, n_1 )$ we consider both possibilities together, and have $P(n_1, n_{Xe}, n_2) + P(n_2, n_{Xe}, n_1 )$.
In general, grouping equivalent partitions favors cases where many equivalent distributions exist over cases where the partitions are unique unique ones (for example, e.g. twelve partitions 1|2|2|3, 1|2|3|2, \dots over 2|2|2|2 for the CH bonds in \ch{CH4}).
A possible limitation of such a procedure is seen when we consider perturbations of the system (chemical reactions): equivalence disappears, grouping is abandoned, and abrupt changes can appear.
\end{enumerate}
For our example, \ch{XeF2}, the distinction between the two variants above is irrelevant; with, or without grouping, the same partitions are selected.

\section{Conclusions}

In 1933, Lewis called his approach analytical:~\cite{Lew-33} deducing from a large body of experimental material some simple laws consistent with the known phenomena.
He contrasted it with the approach of the mathematical physicist who uses postulates to synthesize a molecule.
He recognized its power of producing quantitative results (in contrast to the analytical method that produces only qualitative results).
However, he feared that the inaccuracy of a single postulate might completely invalidate the results while the ``results of the analytical method can never be far wrong, resting as they do upon so numerous experimental supports''.
After one century, the postulates of quantum mechanics not only remained valid, they also allowed to go beyond what had been experimentally known.
However, our minds still aspire for simple laws, in spite of computations becoming more convenient.
This paper tries to extract some simple information from quite accurate computations without trying to explore the origins of the effects observed: we now that they lie in the Schr\"odinger equation and the Pauli principle.

As the core of quantum mechanics is probabilistic, we try to provide probabilistic measures.
We define the space into regions, and compute the probability to have a given number of electrons in it.

We note that the probabilities are low: already the probability to have core-valence separation in \ch{CH4} is only of 0.8.
Considering an arrangement as in a Lewis structure (for example, having two electron pairs in each of the CH bonds) is so low (0.2) that it is more likely to have electrons not distributed according the Lewis structure.
A simple explanation for it is that there are many ways for an electron to quit its pair.
We would like to stress that this is not contradicting the Pauli principle, because we consider electrons in a region of space, not in an infinitesimal volume element.
Not always can we see it as a contribution of ionic structures: such effects show up also between lone pairs on the same atom, or multiple bonds.

Quantum mechanical effects enhance some of the probabilities, and reduce others.
However, such changes are normally around 0.1-0.2.
Such enhancements can be quantified by comparing with models (such as the Ehrenfest urn model), or by considering conditional probabilities.

We try to defend the idea that there is a need to giving the user some freedom in the definitions allowing to adapt the tool to the purpose of the analysis (for example to choose if it is better to refer to atoms or to ions).
Not all interpretative methods give it.
The probabilistic nature provides several interpretations, and depending on what we ask this may be simple or complicated.
We tentatively suggest - without exploring it systematically - an Occam's razor approach inspired from political analyses: the effective number of parties, the parties corresponding to the different distributions of electrons over the spatial regions.
We choose among the partitions that are having the minimal number of representative distributions.

This paper could have had the title: ``Concepts in search for observations''.
It is not clear to us how relevant our remarks are for chemists.
However, these effects are present, and it might be worth looking for them.
To conclude, let us mention an aspect that we could treat with our approach (but have only mentioned in passing in this paper).
One may have regions of space where electrons are localized, producing spatially disconnected regions between which electrons are delocalized.

\clearpage
\appendix

\section{Technical details}
\label{app:technical}
\begin{table}[h]
    \centering
    \caption{Energies (a.u) of the Hartree-Fock and CI wave functions used to generate the configurations.
    $N_\text{det}$ is the number of determinants, and exFCI is an estimate of the full CI energy obtained by
    extrapolation of the CIPSI energies. The percentage of correlation energy are computed with respect to the exFCI energy.}
    \label{tab:data_wf}
    \begin{ruledtabular}
    \begin{tabular}{lcccccc}
                &   \ch{FHF-}  & \ch{KrF2}    &    \ch{XeF2}      &     \ch{F2}      & \ch{HCCH}  & \ch{CH4} \\
      \hline
$N_\text{det}$  & 1083789      &  554150        &  49406              &   260758      &  38038    &  728533 \\
Frozen MOs      &   1--2       &      1--11     &     1--20           &    1--2       &   1--2    &  1      \\
Active MOs      &   3--89      &      12--137   &     21--321         &    3--80      &   3--91   &  2--73  \\
      \hline
HF              & -199.586     &  -2950.704     &  -7430.097    &   -198.765    & -76.851    & -40.213 \\
exFCI           & -200.161     &  -2951.587     &  -7431.468    &   -199.311    & -77.186    & -40.430 \\
      \hline
CI              & -200.138     &  -2951.486     &  -7431.322    &   -199.289    & -77.148    & -40.427 \\
\% Correlation  & 96\%  &  89\% &   89\%  & 96\% & 87\% & 98\% \\
    \end{tabular}
    \end{ruledtabular}
\end{table}

For all the systems presented, we generated a correlated wave function
with a frozen-core configuration interaction using a perturbative selection made iteratively (CIPSI) in the full valence CI space.
For CIPSI calculations, we used natural orbitals of a preliminary CIPSI calculation made with the Hartree-Fock orbitals to generate a more compact CI expansion.
For \ch{XeF2}, the Universal Gaussian Basis Set (UGBS) was used,~\cite{DecJor-98} and for all other compounds we used  the def2-TZVPD basis set.~\cite{SchHubAhl-94}
The CIPSI calculations were performed with the Quantum Package software~\cite{GarAppGas-19} and the QMC calculations were made without any modification of the wave function with QMC=Chem.~\cite{SceCafOse-13}
The energies of the generated wave functions are given in table~\ref{tab:data_wf}.

\section{Smooth boundaries for two particles in a box}
\label{sec:smooth}

\begin{figure}[h]
    \centering
    \includegraphics{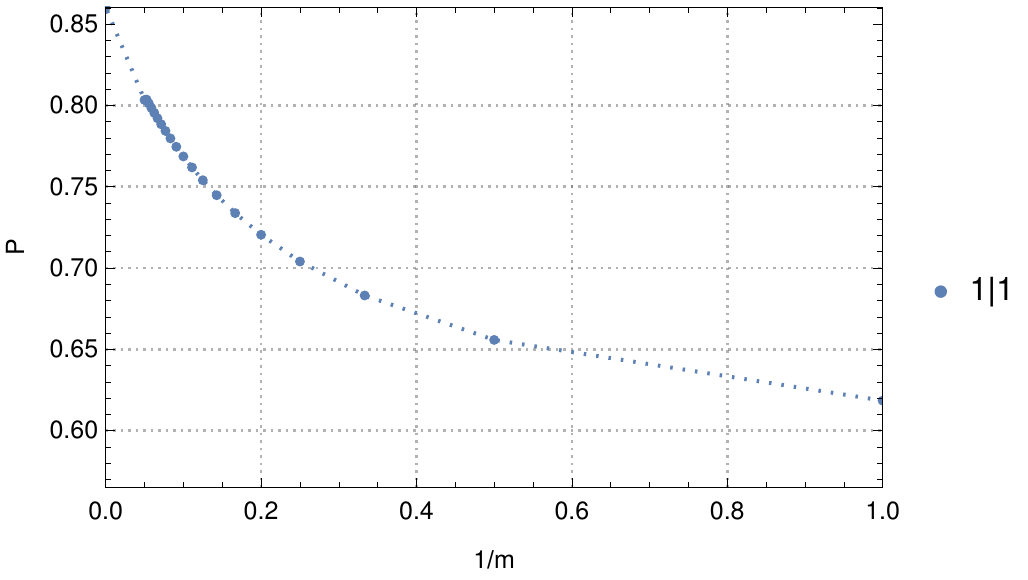}
    \caption{Probability of having one electron in dominantly in the left half of the box (and another electron dominantly in the right half of the box) as a function of the smoothing (using the smoothstep function $S_k$, Eq.~\eqref{eq:smoothstep}). The dotted line is connecting the points to guide the eye.}
 \label{fig:smooth}
\end{figure}

Let us consider two same-spin fermions in a box of length $L=1$.
The eigenfunctions corresponding to the lowest energies are
\begin{align}
    \phi_1(x) & = \sqrt{2} \, \sin( \pi x) \\
    \phi_2(x) & = \sqrt{2}\, \sin(2 \pi x)
\end{align}
yielding for the two particle wave function:
\begin{equation*}
    \Phi(x_1,x_2) = \phi_1(x_1) \phi_2(x_2) -\phi_1(x_2) \phi_2(x_1)
\end{equation*}
We can compute analytically the probability to have one and only one electron in the left (or right) half of the box:
\begin{equation*}
    P_{\text{left}}(1) = P_{\text{right}}(1) = \frac{1}{2} + \frac{32}{9 \pi^2} \approx 0.86, \text{ with sharp cutoff.}
\end{equation*}

Let us now smooth out the indicator function, i.e., use a function that does not define the region by a step function, but by a smoothed step function.
Here we use the smoothstep function,
\begin{equation}
\label{eq:smoothstep}
    S_k (x) = x^{k+1} \sum_{j=0}^k \frac{(k+j)!}{k!\, j!} \frac{(2k+1)!}{(k-j)!\,(k+j+1)!} (-x)^j
\end{equation}
In the limit $k \rightarrow \infty$, it becomes the step function at $x=1/2$, while $S_0=x$.
The expression for the probability to have one and only one particle in one of these ``fuzzy half-boxes'':
\begin{equation}
    P_{left}(1) = P_{right}(1) =  \langle \Phi | S_k(x_1) \left( 1 -S_k(x_2) \right) +
   \left( 1 -S_k(x_1) \right)  S_k(x_2)
   | \Phi \rangle
\end{equation}
Fig.~\ref{fig:smooth} shows the change of the probability with the change of the smoothness parameter.
(Note that the points in Fig.~\ref{fig:smooth} show $P_{\text{left}}$  as a function of $1/k$: the values at the left of the graph correspond to a strong cutoff.)
The highest value of the probability is reached for the sharpest cutoffs, decaying toward
\begin{equation*}
    P_{\text{left}}(1) = P_{\text{right}}(1) = \frac{1}{2} + \frac{512}{81 \pi^4} \approx 0.56, \text{ with }S_0(x)
\end{equation*}
Thus, replacing the indicator function by a smoothstep function has an equalizing effect on the probabilities.
The probability to have any of the ``ionic'' structures, i.e., two electrons in one half of the box, and no electrons in the other is given by
\[  \frac{1}{2} - \frac{32}{9 \pi^2} \]
for the sharp separation.
This value increases as the indicator function is smoothed out, giving more weight to the minority distribution as the $k$ increases.

\clearpage
\bibliography{biblio-proba}{}
\end{document}